# Investigating differences in lab-quality and remote recording methods with dynamic acoustic measures

Cong Zhang[1], Kathleen Jepson[2], Yu-Ying Chuang[3]

[1]Newcastle University

[2]Institute of Phonetics and Speech Processing, Ludwig-Maximilians-Universität Munich

[3]Eberhard Karls Universität Tübingen

Corresponding author: Cong Zhang (cong.zhang@newcastle.ac.uk)

**Abstract**

Increasingly, phonetic research utilizes data collected from participants who record themselves on readily available devices. Though such recordings are convenient, their suitability for acoustic analysis remains an open question, especially regarding how the individual methods affect acoustic measures over time. We used Quantile Generalized Additive Mixed Models (QGAMMs) to analyze measures of F0, intensity, and the first and second formants, comparing files recorded using a laboratory-standard recording method (Zoom H6 Recorder with an external microphone), to three remote recording methods, (1) the Awesome Voice Recorder application on a smartphone (AVR), (2) the Zoom meeting application with default settings (Zoom-default), and (3) the Zoom meeting application with the "Turn on Original Sound" setting (Zoom-raw). A linear temporal alignment issue was observed for the Zoom methods over the course of the long, recording session files. However, the difference was not significant for utterance-length files. F0 was reliably measured using all methods. Intensity and formants presented non-linear differences across methods that could not be corrected for simply. Overall, the AVR files were most similar to the H6's, and so AVR is deemed to be a more reliable recording method than either Zoom-default or Zoom-raw.

## 1. Introduction

The feasibility of analyzing speech data collected by alternative recording methods, outside of the laboratory, with accessible equipment, has been of interest to researchers for some time. This line of research has been driven by a range of needs, including access to particular participant groups (De Decker & Nycz, 2011; Vogel et al., 2015), frequent data collection for clinical voice assessment (Grillo et al., 2016), use of existing speech data for research purposes (Bulgin et al., 2010; Fuchs & Maxwell, 2016; Rathcke et al., 2017), and the possibility of running small studies without using limited laboratory resources, e.g. undergraduate dissertation projects. The broader need for better understanding of the effects of alternative recording methods was highlighted during COVID-19-related measures. Restrictions on travel and in-person meetings resulted in laboratory-based production studies needing to take a very different form than before if they were to be carried out. While people are adapting to a more hybrid norm, the pandemic continues to influence the way we live and work, and we realize as a field that financial, logistical, ethical, and political issues will remain a barrier to conducting laboratory or in-person data collection. Therefore, there is a need for further investigation into the comparability of laboratory-style data to data collected by alternative methods, particularly those that put the recording device into the hands of participants.

One of the commonalities in previous research is that it relies on mostly static measures, such as mean F0 (e.g. Uloza et al., 2015; Zhang et al., 2020, 2021), and steady state measurements of formants (e.g. Sanker et al., 2021; Zhang et al., 2021). The existing research, reviewed in more detail below (Section 1.1), cannot address the needs of many researchers, such as researchers in prosody who are interested mostly in dynamic measures of F0 and intensity, for example. The need for dynamic measures is generally being recognized within phonetics research – Sóskuthy (2021) defined "dynamic speech analysis" as "the analysis of phonetic contours", which can be either temporally or spatially ordered measurements. A second shared feature of previous studies is that they mostly examined speech materials that were short, such as read words (Freeman & De Decker, 2021b), sustained vowels of 2-5 seconds in duration (e.g. Ge et al., 2021; Zhang et al., 2021), and synthesized sounds (e.g. Manfredi et al., 2017). Isolated words and connected speech differ in part due to various connected speech phenomena (e.g. Harmegnies & Poch-Olivé, 1992), such as vowel reduction, and can therefore result in a shift in how discovered differences are interpreted. A small number of studies looked at longer materials (e.g. Fahed et al., 2022; Jannetts et al., 2019; Maryn et al., 2017; Penney et al., 2021; Vogel et al., 2015); however, they did not examine dynamic measures, and indeed, only examined single tokens extracted from these longer recordings. Considering both points, by examining dynamic measures of connected speech, we would be able to reveal differences between recording methods which may be otherwise hidden behind the mean values.

## 1.1. Background

Previous studies have investigated the effect of different hardware, file formats, and software on a range of speech measures using different types of speech materials.[1] Appendix 1 summarizes studies of alternative recording methods in a comparable formant to Jannetts et al. (2019), with details of the speech data analyzed, hardware, file formats, and software evaluated. As shown in Appendix 1, a diverse range of parameters have been examined in the literature, including signal-to-noise ratio (Kojima et al., 2018; Maryn et al., 2017) and voice quality measures such as jitter, shimmer, cepstral peak prominence (CPP) and harmonic-to-noise ratio (HNR) (Fahed et al., 2022; Grillo et al., 2016; Kojima et al., 2018; Maryn et al., 2017; Uloza et al., 2015; Vogel et al., 2015). Importantly, recent studies, which use the most contemporary devices and software options, conclude with advising caution when measuring many of these voice quality and amplitude-based measures (e.g., HNR, shimmer, etc.) (Fahed et al., 2022; Penney et al., 2021). However, it is noted that, along with device- or software-based issues, there may be a number of contributing factors to the unreliability of remote methods, for example, being unable to control for environmental noise, the position of the device, or the microphone quality (Fahed et al., 2022), which are inevitable in the context of at-home recordings even with mitigating actions.

In the remainder of this section, we focus on reported findings for the measures of interest in this paper, namely F0, formants, and intensity, as well as temporal differences caused by different recording methods.

Many previous studies looked at measures of F0. Together they showed that F0 is a robust measure that is accurately captured by a range of devices and file formats (see table in Appendix 1). When F0 differences are reported between baseline and comparison methods, they are often very small (e.g., Fahed et al., 2022 report a difference of 0.66 Hz between baseline method and smartphone condition, and –0.53 Hz between baseline and tablet condition for participants with Huntington's disease; though we note there was no significant difference between the baseline and test methods for a neurotypical control group). In most studies, the mean values of F0 of sustained vowels is measured and compared (Fahed et al., 2022; e.g. Uloza et al., 2015; Vogel et al., 2015; Zhang et al., 2021); some researchers have examined F0 in read stories (Fahed et al., 2022; Jannetts et al., 2019; Maryn et al., 2017; Penney et al., 2021; Vogel et al., 2015). In all instances, however, what was measured was single F0 values of vowels, such as mean, median, and standard deviation. One problem with these measurements is that they do not capture subtle F0 modulations that are important in many studies, such as studies of intonation or word-level prosody, and it is not possible to extrapolate from steady-state F0 to the F0 of utterances. The current study extends what we know about the reliability of F0, considering F0 contours at the utterance level.[2]

---

[1] In this section we refer to software options for making audio recordings. These include conferencing software – Zoom (https://zoom.us/), Skype (https://www.skype.com/en/), Microsoft Teams (https://www.microsoft.com/en-us/microsoft-teams/group-chat-software); smartphone messaging software – Messenger (https://www.messenger.com/); software for audio interviews – Cleanfeed (https://cleanfeed.net/); and software for audio recording and analysis – Praat (Boersma & Weenink, 2020).

While not under investigation in the current study, we note that F0 is found to be affected by extreme file compression. Fuchs and Maxwell (2016) examined the effect of compression rates between 16 and 320 kbps on a range of F0 measures across vowel, obstruent and sonorant segments from words in read speech. They concluded that mp3 data are viable for F0 analysis, when compressed at bit rates between 56 kbps and 320 kbps; however, greater errors were found for more extreme compression rates (e.g., lower bit rates such as 16 kbps and 32 kbps).

Unlike F0, formants have presented a more complicated case, as they have been found to not only be affected by recording method and file type, but also these interact with speaker gender, and further, individual vowels are affected differently. For example, Zhang et al. (2021) examined mean F1, F2, and F3 of sustained vowels, and concluded that smartphone recordings captured the formants more accurately than the online conferencing software Zoom (henceforth *Zoom*): the smartphone recordings did not present a significant difference from the baseline, recorded with a Zoom H6 digital recorder (henceforth *H6*). The Zoom conferencing software performed poorly in capturing all three formants. They also observed more errors in the Zoom recordings from female speakers than from male speakers. De Decker and Nycz (2011) also observed a gender difference. In their study, they examined vowels in h_d contexts spoken by one female and one male speaker and compared an iPhone (recording m4a), Macbook Pro (recording wav in Praat), a Mino Flip video camera (avi converted to aiff), and YouTube audio (downloaded as mpa) with recordings made using a Roland Edirol recorder. They reported that the lossy avi files from the camera had higher F1 values for both speakers than the baseline Edirol wav recordings. However, the effect was stronger for low back vowels of the male speaker and the high vowels of the female speaker. For F2, on the other hand, the male speaker's vowels were not affected by recording type, but the female's front vowel measurements were higher and those of the back vowels lower, resulting in a distortion of the vowel space (De Decker & Nycz, 2011). Further, the compression used in the transmission of speech data via Skype has been found to significantly alter formants such that vowel spaces may be expanded in both F1 and F2 dimensions, or the vowel space is distorted with expansion in one part and compression in another (Bulgin et al., 2010). A more recent study reported that Skype, Zoom, and Microsoft Teams (henceforth *Teams*) faithfully maintained patterns of overall vowel spaces for both a female and a male speaker but showed deviation of absolute formant values in the range of 750-1500 Hz, resulting in specific issues for mid-back vowels (Freeman & De Decker, 2021b). Freeman and De Decker (2021b), who examined vowels in read word lists, reported that Teams was most accurate for the female speaker while for the male speaker the low, back part of the vowel space was compressed. Skype, on the other hand, was most accurate for the male speaker but least accurate for the female, expanding the vowel space except in the area of the high front vowels. It is noted that the effects of gender are not clear at this stage because some of these studies include only one speaker of each gender, and so observed differences may be due to individual participants. In the current study, we investigated F1 and F2 of vowels, and examined the effects on the vowel space using three representative corner vowels, also considering speaker gender.

---

[2] See Supplementary Material 3 for an analysis of F0 over vowels.

Intensity of the speech signal overall is not frequently investigated (c.f. Sanker et al., 2021). One reason for this is that the distance from the microphone can result in differences in absolute intensity (Sanker et al., 2021), making comparisons of single time point or mean measures uninformative. Sanker et al. (2021) found differences in intensity in the test recording conditions compared with their baseline Zoom H4N recorder, as expected. Sanker et al. (2021) and Penney et al. (2021) also investigated a range of voice quality measures including spectral tilt (e.g., H1-H2), which rely on accurately measuring the amplitude of frequencies within the speech signal. Spectral tilt was found to be affected by the software in Sanker et al. (2021) with lower values for all software options, except for Messenger, which was higher, and to a greater degree than any other tested software method. Regarding device, only their Android method resulted in significantly different spectral tilt values from their baseline method, once again, with a lower value. Penney et al. (2021) found that H1 was significantly higher across their test devices compared with their reference level H6 recorder. As a reviewer pointed out, microphone sensitivity to various frequencies in addition to any processing by software programs, such as those suggested by Sanker et al.'s (2021) results, could affect the relative intensity at different frequencies, and this could affect intensity over an utterance because of the different distribution of frequencies throughout a word or utterance. Zhang et al. (2021) observed unexpected periods of extremely reduced intensity for Zoom recordings with default settings. Overall, these results suggested a need for the investigation of the intensity measure, and so we included the examination of intensity contours in our current study.

It is not yet known how speech is affected over time by alternative recording methods. Sanker et al. (2021) reported issues with alignment of long files in recordings made with the Zoom conferencing software, Cleanfeed, and Messenger. Ge et al. (2021) likewise reported that cloud-based Zoom recordings were significantly different than the Zoom H2N baseline with respect to the duration of some speech sounds. Fricatives, for example, were 19.83 ms shorter in Zoom recordings than in baseline recordings, while vowels were shorter by 7.51 ms on average; the other segment types were comparable (~ 1 ms). Ge et al.'s (2021) findings may reflect a difference in the Zoom software, but it may also reflect a difference in the visual cues used in manual segmentation, which have been found in inter-rater reliability studies to be in the region of 10 – 20 ms (see Machač & Skarnitzl, 2009, pp. 13–14 inter alia). We therefore consider duration differences with respect to the segmentation process, and the temporal alignment of landmarks across recordings.

## 1.2. *Current study*

In this study, we present results from a comparison of simultaneous recordings made using four recording methods; a baseline recording method which is representative of laboratory settings and three recording methods that could be broadly accessible to people at home. The baseline method is a high-quality digital recorder (Zoom H6 recorder, henceforth *H6*), and the three remote methods are: (1) a smartphone with a non-lossy wav formant recording application, Awesome Voice Recorder (Newkline, 2020) (henceforth *AVR*); (2) a computer running the conferencing software Zoom with default post-processing (e.g., noise cancelling) enabled (henceforth *Zoom-default*); and (3) a computer running Zoom without post-processing (henceforth *Zoom-raw*; see section 2.3 for a full description).

The two comparison applications (AVR and Zoom) can be used on a wide range of hardware options to allow flexibility for participants. The focus of this paper is to evaluate and compare these specific applications, not smartphone or laptop makes or models. AVR was selected because it records non-lossy wav mono-channel files with high sampling rate, is available on iOS and Android, and saves recorded data locally. Zoom was selected because it is one of the most commonly used online conferencing apps and provides two recording options: one offers a noise cancelling function that reduces background noise to produce clearer audio without use of professional headphones and microphones; the other provides a possibility to retain the raw features of the original recording without post-processing. These "at home" recording methods were selected to represent some of the ways that participants in speech production studies could record at home without additional equipment, though it is acknowledged that there are many other methods using combinations of software, devices, and additional equipment that could be tested. More importantly, all the methods in the current study can be operated without the need for using a cloud server to store the data, which allows the data collection to follow the data security guidelines of many funding organizations.

This study is motivated by the need to understand the effects over the course of larger units, such as utterances as these are highly relevant for prosodic analysis. With the dynamic approach taken in this paper, we can examine differences in F0, intensity, and formants over time in curve height and curve shape, allowing for comparisons that go beyond mean or single time point differences. We have also been prompted by a general concern that convenient recording options employ unknown methods to modify audio, primarily to remove unwanted noise. It is not clear how this affects audio recordings over time, with the methods potentially affecting some parts of the audio while not others. Therefore, we compared baseline recordings (H6) with three test conditions (AVR, Zoom-default, and Zoom-raw) to make direct recommendations based on these specific computer and smartphone applications. Regarding analysis, this study examines F0 over utterance, intensity over utterances, as well as vowel F1 and F2 to assess overall effects of recording method on these regularly examined measures. Two temporal analyses were also performed: a comparison of utterance duration, and a comparison of landmark time points across entire recording session files. While we cannot say for certain why some of these results occur, we put forward some speculations.

## 2. Materials and methods

### 2.1. Participants

Eight speakers, four female (PF1-PF4) and four male (PM1-PM4), aged 28 to 32 years (mean = 30.4, SD = 1.4) took part in the study. Speakers PM2 and PF3 were monolingual speakers of Australian English, PF4 was a multilingual speaker with American English as her first language, PF1 and PF2 were multilingual with Mandarin as their first language, and PM1, PM3, and PM4 were multilingual with Dutch as their first language. The variable linguistic backgrounds of the participants are not a problem for the present study which focuses on differences between recording methods. Participants were recruited from Radboud University, and were aware of the purpose of the study. Five of the participants had linguistic training.

We acknowledge the number of participants recruited for this study is small. Due to COVID-19 restrictions, the authors' institutions were not accessible for recordings or larger-scale recruitment at the time of data collection in January 2021, and in-person data collection was not possible due to limitations on gatherings in private settings. However, the number of participants is comparable to similar studies of the same nature (c.f. studies in Appendix 1). Appendix 1

### 2.2. Materials

The test materials were five sonorant-rich, scripted utterances that varied in their intonational tunes. They were designed to include rises, falls, and level stretches of pitch. Demonstration audio files were recorded for these utterances using varied liveliness levels which differed in loudness and pitch dynamism. Examples (1)-(5) (henceforth, Utterances 1-5) provide illustrations of the demonstration files. The speakers were only presented with the text and audio without any visualizations (see further details in 2.3 and 2.4); however, here we present the visualizations for the reader. Contextual information is provided in brackets, beneath which is the prompt text; the blue lines display F0 (Hz) and the red lines display intensity (dB). The utterances were also designed to contain a range of English monophthongs and diphthongs. Measurements from both the utterances and a selection of vowels were analyzed (see Section 2.6 for details).

(1)    (Is this even food?! It's inedible!)
       My ramen aren't inedible!

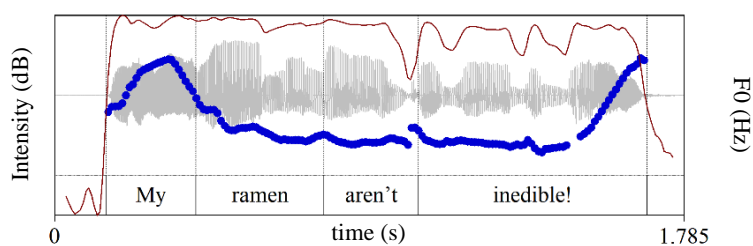

(2)    (What did Emmanuel make for the bake sale?)
       Emmanuel made the banana bread.

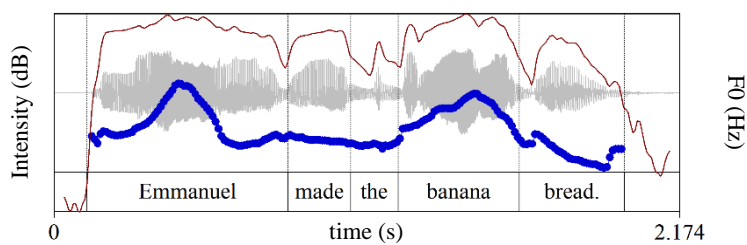

(3)     (Take as many mangoes as you want! I've got a free supply for a year!)
        Free mangoes for a whole year?!

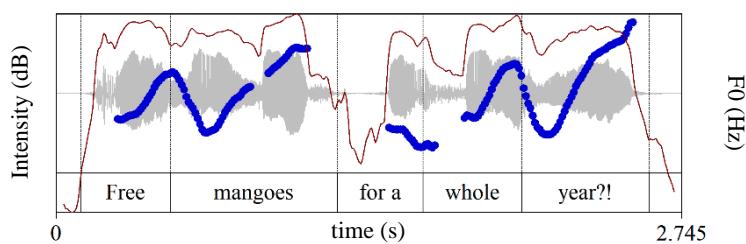

(4)     (*Calling someone*)
        Amelia! Your noodles are ready!

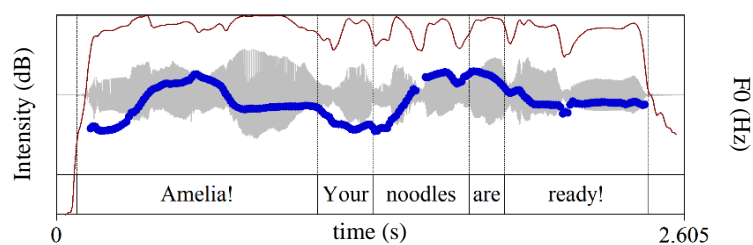

(5)     (Did you eat your stew?)
        Do you mean my goulash? It's a soup you know.

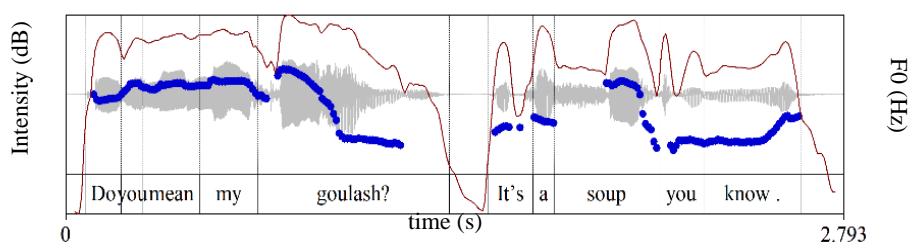

## 2.3.    Recording equipment and set-up

To be able to conduct this study, we required that all participants used the same type of high-quality recording device as a baseline for comparison; therefore, it was necessary for participants to have access to an H6 with headset microphone. We provided these to each participant. Participants made use of their own computers and smartphones for the alternative recording methods, detailed below.

Participants recorded themselves, following the protocol in Supplementary Material 1, with a Zoom H6 recorder with Sennheiser HSP2 headset microphone, a personal smartphone to run the AVR recording application, as well as two personal laptop computers to run the two Zoom conditions simultaneously. Smartphones included both iPhones and Android devices, and computers included devices running both MacOS and Windows OS; see Table 1 for detailed information. Differences between smartphone models and computer models are not investigated here since it was not our goal, and it would be impossible to control the personal devices participants owned. Different smartphones and computers can have an effect on some measures, specifically, sensitive voice quality measures (Jannetts et al., 2019; Penney et al., 2021), but using different equipment does not always play an important role for all measurements (e.g. Jannetts et al., 2019; Zhang et al., 2021). In this study, there was considerable overlap between participants and devices, meaning that it would be difficult to assess what was the effect of device and what was attributable to the individual participant. In the statistical models, participants were accounted for so including equipment would create a confounding factor. Given that the phone/computer models were all randomly distributed across recording devices (i.e., not only one model for one recording device), the effect of device was unlikely to be crucially determined by equipment.[3] Future studies are indeed needed for investigating further into the differences between equipment options.

Table 1: Recording equipment for each participant

| Speaker ID | AVR phone model | Zoom-default computer model | Zoom-raw computer model |
|---|---|---|---|
| PF1 | iPhone 8 | Mac Book Pro 2014 | Acer aspire 5600 |
| PF2 | Samsung Note 10 | Microsoft Surface Pro 6 | Microsoft Surface Pro 6 |
| PF3 | Google Pixel 3a | Lenovo Thinkpad T495 | ASUS UX330U Notebook |
| PF4 | iPhone 8 | Mac Book Pro 2015 | Mac Book Pro 2016 |
| PM1 | iPhone 12 | Mac Book Pro 2015 | Mac Book Pro 2016 |
| PM2 | Google Pixel 3a | ASUS UX330U Notebook | Lenovo Thinkpad T495 |
| PM3 | iPhone 8 | Mac Book Pro 2015 | Mac Book Pro 2016 |
| PM4 | One plus 6 | Mac Book Pro 2015 | Acer aspire 5601 |

---

[3] The results also did not appear to have varied systematically by equipment.

Participants were provided with a PowerPoint presentation with detailed instructions for setting up the devices. We provide the full instruction document in Supplementary Material 1 for the convenience of future researchers and educators who wish to use one of the methods reported in this paper. The instructions included how to update settings for software, and prepare the H6 for recording (see below), as well as what to do during the recording (e.g., clap before each utterance, imitate recordings as fluently as possible), and after (e.g., how to save and send their files). Recording set-up instructions were provided pictorially to participants using Figure 1: the smartphone with AVR was to be placed on a soft material such as a towel directly in front of the participant at a distance of 20-30 cm. The microphone at the bottom of the smartphone was to be pointed at the participant. The Zoom-default and Zoom-raw meeting computers were to be placed directly in front of the participant, approximately 40-50 cm away, resembling a Zoom meeting setup. The Zoom-default computer was also used for displaying the PowerPoint presentation that contained the study instructions and prompts. The H6 was used with a head-mounted microphone. Participants adjusted input levels and were instructed to aim for a maximum input level in their normal speech of -12 dB to avoid clipping; levels could be monitored on the H6 device screen before recording and input levels could be adjusted using the level dials.

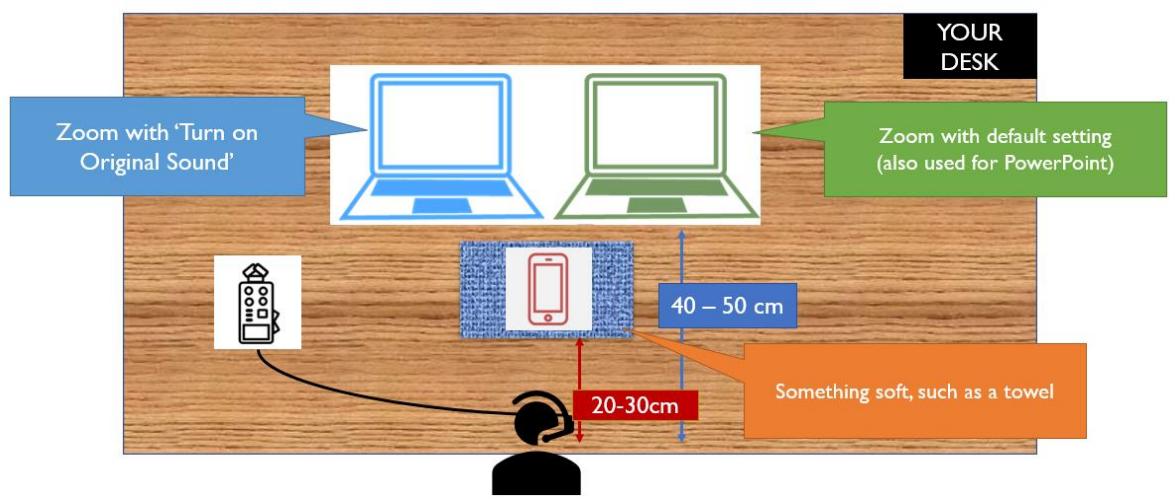

Figure 1. The set-up of H6, smartphone with AVR, Zoom-default computer and Zoom-raw computer.

H6 was set to record mono-channel wav files, at 44.1 kHz, 24 bits. AVR was set to record mono wav files, at 44.1 kHz, 256 bps. Zoom Version 5.4.9 (59931.0110) was used for both Zoom conditions, which recorded stereo-channel m4a files. For Zoom-default, the default Zoom settings with post-processing (e.g. noise-cancelling) were used. For Zoom-raw, the "Turn on Original Sound" setting was used, along with "Disable echo cancellation" and "High fidelity music mode". The advanced option of "Signal processing by Windows audio device drivers" on Windows computers was set to "Off (Windows – Raw)". These options should allow Zoom to record with as high the fidelity as possible, without using standardly applied Zoom audio filters or sound altering features. Both Zoom computers were connected to the internet; however, the participant was the only person in the Zoom meeting, and recorded themselves using the "Record" function so the recording quality was not affected by internet connection. Computer internal microphones were used for both Zoom conditions. Zoom m4a files were converted to mono-channel wav files at 44.1 kHz and 256 bps using VLC (VideoLan, 2019).

### 2.4. Recording procedure

All recordings were made in quiet locations in the participants' homes, where environmental noises were limited as much as possible. Speech data were simultaneously recorded using all four recording methods. Participants were asked to turn all devices to silent mode. Both computers' speakers were turned off to avoid feedback. Participants were not asked to restart their devices or stop all other processes on their devices before recording. This was not possible because, as mentioned, participants viewed the PowerPoint prompt on the Zoom-default device. In this way, the Zoom and AVR recordings reflected a real-world use of the devices in a remote recording setting in which participants may be required to view files on their recording device in order to read target utterances or texts, describe prompt images, or play elicitation games.

Along with recording instructions, the PowerPoint presentation mentioned above in Supplementary Material 1 also contained the speech materials and recording procedures. Participants were presented with each utterance (see 1–5 in 2.2) orthographically on a separate slide three times in pseudo-randomized order. A sentence providing contextual information was also provided on the slide, along with the demonstration audio file of the target utterance for participants to imitate. Imitation was used as we wanted to elicit similar contours across participants, and pitch tracks with extensive pitch excursion. Participants were asked to clap once at the start of the recording session. For each utterance repetition, the participant played the illustrative audio file from the PowerPoint and then clapped their hands, paused for approximately 1 second, produced the utterance, and paused before proceeding to the next slide. This procedure of listening to the audio, clapping, and then speaking was used for each utterance and repetition. The claps were used to demarcate the onset of each utterance and to examine duration differences throughout the full recording session audio files. At the end of the task, each participant saved the recordings from each device (four files total) which contained all utterance repetitions. Participants emailed these files to the researchers along with a metadata document containing information about the participant, and the hardware they used to record.

*2.5.  Data processing*

Each utterance repetition from the recording methods was segmented as an individual file, including the preceding clap which was used to align the signal across matched files. This resulted in a total of 480 utterances for analysis (5 utterances × 8 participants × 3 repetitions × 4 methods).

During data preparation, it was noted that there was a temporal difference between files such that over the course of the long, recording session files, the clap landmark points between the H6 and comparison methods diverged over time. We determined that this could be a relevant issue; an analysis of temporal aligned was incorporated into the study, and is discussed in Sections 2.5.1, 2.6.1, and 3.1.

Eleven phonemic vowels /iː, ɪ, e, ə, æ, ɑː, uː, aɪ, eɪ, iə, əʊ/ were selected for the combined vowel analysis to assess if methods had an overall effect on vowel formants (see Section 2.62.6.2). These vowels were selected because all speakers produced them and in a relatively consistent way. It was also possible to reliably segment them from the neighboring segments. The vowel categories were identified in 26 words from the five utterances, resulting in 26 target vowels which were unique in terms of their vowel category, word context, and utterance context (see bold orthographic vowels in Table 2.  A total of 2496 vowels were analyzed in the combined vowel analysis (26 target vowels × 8 participants × 3 repetitions × 4 methods).

For the vowel space analysis, the three vowels /iː, æ, uː/ were selected as they represent the high-front, low, and high-back points of the vowel space. These occurred in nine words from the utterances, resulting in nine target vowels (see bold vowels in grey rows in Table 2). A total of 864 vowels were analyzed in the vowel space analysis (9 target vowels × 8 participants × 3 repetitions × 4 methods). The low-back vowel /ɑː/ was excluded from the vowel space analysis because it did not occur as low-back in the speech of all speakers since the native languages and varieties of English spoken by the speakers did not contain /ɑː/.

Table 2: Vowel categories and their occurrences in words

| Vowel | Word | Standard lexical set keywords |
|---|---|---|
| iː | Amelia, free, mean | FLEECE |
| ɪ | Emmanuel, inedible | KIT |
| e | bread, ready, inedible | DRESS |
| ə | banana, ramen, Amelia | commA |
| æ | Emmanuel, goulash, mangoes | TRAP |
| ɑː | aren't, banana, ramen | START |
| uː | goulash, noodles, soup | GOOSE |
| aɪ | my (Utterance 1), my (Utterance 5) | PRICE |
| eɪ | made | FACE |
| iə | Amelia | NEAR |
| əʊ | know, mangoes | GOAT |

Vowels are well known to vary considerable across English varieties (e.g. Clopper et al., 2005; Cox & Palethorpe, 2007; Wells, 1982). There are also differences in the pronunciation of some of the above words. For example, "goulash" may have either /ɑː/ or /æ/ in the second syllable, and the final vowel in "Amelia" may be realized as a diphthong or as two syllables, depending on the variety of English. These differences do not pose an issue for the current analysis in so much as each vowel is compared across recording devices, for each speaker, and each repetition. They could, however, present challenges in interpreting the results of the detailed analysis of the three vowels /iː, æ, uː/. These vowels showed variability between participants, attributable to the different English varieties spoken, as can be observed in Figure 2 which plots mean raw Hz values for F1 and F2 of these vowels for each token from the four recording methods. We can see in particular, for the two Australian English speakers (PF3 and PM2), a bimodal distribution of the /uː/ vowel, due to the fronting of the vowel in "noodles" and "soup", and a more back realization in "goulash". We can say that, for all speakers, while there is some overlap between the categories, they remain broadly representative of corner vowels in that they are three vowels maximally distributed in the vowel space. Therefore, we deem that they are suitable for the illustrative examination of the recording methods, below, and can be extrapolated to the analysis of other vowels.

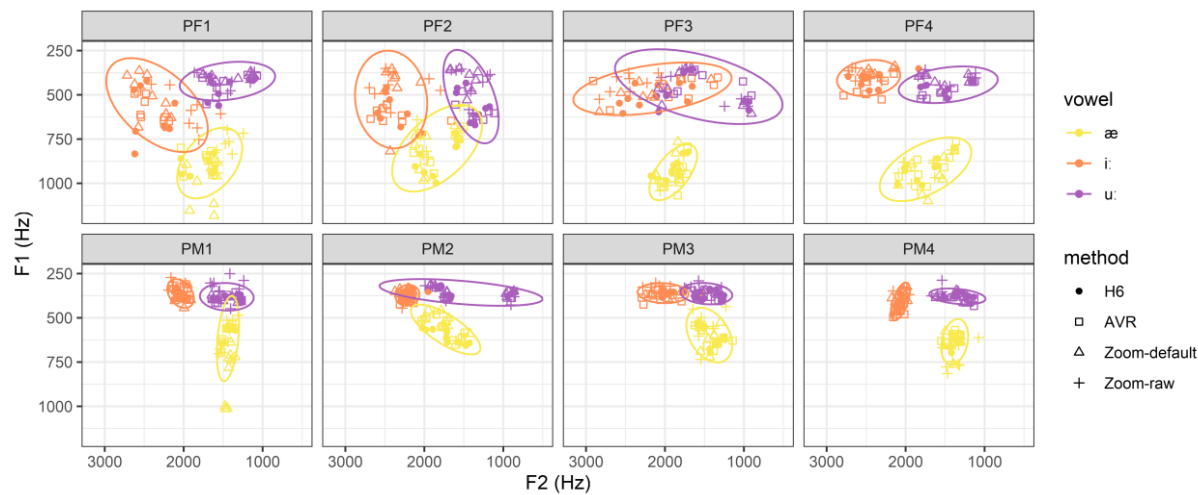

Figure 2: Mean raw F1 and F2 values (Hz) for each token of TRAP (/æ/ "Emm**a**nuel, goul**a**sh, m**a**ngoes") FLEECE (/iː/ "Am**e**lia, fr**ee**, m**ea**n") and GOOSE (/uː/ "**gou**lash, n**oo**dles, s**ou**p") vowels plotted by recording method for each speaker, with ellipses representing 95% confidence intervals.

### 2.5.1. *Praat TextGrid annotation*

The data were annotated in Praat (Boersma & Weenink, 2020). Three sets of TextGrid annotations were used. The first set of annotations was made to investigate temporal alignment differences in the whole recording session files. Recording session files were between ~190 s and ~350 s in duration with each file in a participant's set being approximately similar in duration; for example, PM1's files were 188, 191, 193, 195 seconds in duration – slight variability was due to differences in when the participant commenced recording on each recording device. To be able to compare across files irrespective of these commencement differences, the matched recording session files were combined as individual tracks into one file and aligned to the first clap of the recording session. Three time points in each file set were annotated: 1) the onset of the clap before the first utterance (coded as *early*), 2) the onset of a clap around the middle of the file (coded as *mid*), and 3) the onset of the clap before the last utterance (coded as *end*).

For the analysis of utterance duration, each utterance was saved as an individual wav file, and the onset and offset of each utterance across all four methods were manually annotated in a second set of TextGrids. This was done to compare how different recording methods influence utterance duration and manual segmentation precision.

The last set of annotations were used for the analyses of F0, intensity and formants across utterances and segments. This involved two sets of TextGrids. In the first set of TextGrids the onset clap before the utterance was annotated for each utterance file from the four recording methods. The second set of TextGrids was used to extract the measures and was created as follows. Using the H6 individual utterance wav files and TextGrids, segments were forced aligned using the 'Interval-Align Interval' function in Praat with Language set to British English. Segmentation was manually corrected following the criteria in Machač & Skarnitzl (2009). To use those hand corrected TextGrids to extract acoustic measures across files from the three test recording methods we made duplicate TextGrids with adjusted timestamps. To adjust the timestamps, corresponding individual utterance wav files were aligned using the onset of the clap before each utterance extracted from the above mentioned TextGrids with the annotated claps. In the TextGrids for measurement extraction, the starting time of each utterance from each method was adjusted using the time difference between the claps across different methods. This was to ensure the TextGrids were aligned with the audio from each method. The H6's utterance length was used to ensure the same number of data points could be extracted. In this way, we ensured that we extracted the measures of interest from the same time points across files from the four recording methods.

### 2.5.2. Measure extraction

Praat was used to extract all acoustic measures. For utterances, the measures of F0, and intensity (raw values and normalized) were extracted every 10 ms across entire utterances. Intensity was normalized using the "Scale intensity" function in Praat using the standard value of 70 dB as the new average intensity. Raw intensity values were extracted from the files as recorded. For the vowel analyses, F0, F1, and F2 were extracted every 10 ms based on onset and offset boundaries of the vowels. F0 was extracted using a range of 30-650 Hz after inspecting the minimum and maximum F0 values in the whole corpus. F1 and F2 data were extracted using a range of 0-5000 Hz for male speakers and 0-5500 Hz for female speakers.

### 2.6. Data analysis

### 2.6.1. Utterance duration and temporal alignment difference

As mentioned in Section 2.5, utterance start and end times were manually annotated in the individual utterance files to compare their duration across recording methods. The aim of this comparison was to investigate how reliable the home recordings were in terms of duration, as this has been found to be an issue in files that are compressed/decompressed or transferred over the internet (Sanker et al., 2021). Since the calculation had to inevitably rely on manual segmentation, the duration was also a reflection of how easy and accurate manual segmentation can be for the audio files recorded using different methods. We would therefore alert the reader that any discrepancies included both factors: method-originating temporal differences and differences due to ease of segmentation for a human annotator.

Utterance duration from the baseline H6 was compared with utterance duration from the three comparison methods. A linear mixed effect model was built in R (R Core Team, 2021) using the lme4 package (Bates et al., 2015) to test whether the differences were statistically significant. The dependent variable was DURATION; RECORDING METHOD was the fixed effect, and REPETITION, SUBJECT, and UTTERANCE were fitted as random intercepts.

To investigate temporal alignment over the whole recording session files, simple linear models were fitted for each recording method to examine whether the temporal difference was in a linear relationship with the time points in H6 recordings.

### 2.6.2. Dynamic analysis of F0, intensity, F1 and F2

In order to trace the dynamic difference across different recording methods, we analyzed the acoustic measures (F0 and intensity for utterances; F1, F2 for vowels) using Generalized Additive Mixed Models (GAMMs, Wood, 2017). GAMMs are "an extension of generalized linear mixed models" that as well as parametric terms, allow for the inclusion of smooth terms which model nonlinear shapes, and estimate their degree of wiggliness. Therefore, GAMMs enable us to model nonlinear effects of the predictors, and for this reason they have been used to study response variables in phonetics that vary along the temporal domain, such as tongue movement trajectories (Wieling et al., 2016) and pitch contours (Chuang et al., 2021; Kösling et al., 2013; Sun & Shih, 2021).

GAMMs build on the assumption that the residual errors should be independently and identically distributed. The current dataset, however, consists of a substantial number of extreme values. As these outliers are potentially informative about the recording quality of different methods, we did not remove these data points just to meet the requirement of homoscedasticity. We therefore turned to the quantile GAMMs (QGAMMs, Fasiolo et al., 2020), an extension of GAMMs that makes it possible to model different quantiles in the distribution of the response variables. Importantly, QGAMMs do not have any distribution assumption with regards to residuals, so we could still model the time-varying effect of the acoustic measures while retaining all our data points.

Since we are interested in the differences in recording methods, we fitted QGAMMs, with difference smooths. That is, we set H6 as the reference level, and directly modeled the difference between H6 and the other three recording methods across time. In addition, we included nonlinear by-speaker and by-item (utterance or vowel) random effects by means of factor smooths, so that the method differences that we observed steered away from speaker and item variability. It is worth noting that while GAMMs can deal with the issue of autocorrelation, that is, the response variable at time t is dependent on that at time t-1, this is not yet implemented in QGAMMs. Because of this limitation, and the size of our dataset, we were cautious and remained conservative about the effects, only considering the effects to be significant when the *p*-value was smaller than 0.0001 (c.f. Chuang et al., 2021).

## 3. Results

In reporting the results of the QGAMMs, both contour height and contour shape are discussed. For each analysis, we first present an average plot to illustrate the raw data produced by different methods. Then, the model outputs from the QGAMMs are reported in tables. We compare contours from the three test methods with the contours from the H6. Each table is divided into two parts: parametric coefficients and approximate significance of smooth terms. A significant difference in the parametric coefficients ($Pr(>|z|) < 0.0001$) suggests that the contours, irrespective of their shape, are different in height at the intercept. A significant difference in the approximate significance of smooth terms (p-value $< 0.0001$) suggests that the contours differ in terms of shape, or their trajectory, but does not specify in a particular direction (as it may vary) nor where the difference is observed along the contour. In the smooth term model summary, an effective degree of freedom (edf) indicates the relationship between the two contours, for example, an edf of 1 indicates that the relationship is linear. Lastly, we use the difference plots to illustrate the model results and show how different methods deviate from the H6 baseline method. The difference plots can shed light on where the contours vary. The model estimate of each comparison method is shown as a solid line with $\pm 2$ standard errors in dashed lines. When a recording method is not different from the baseline method, the area between the dotted lines includes the horizontal reference line ($y = 0$). When there is a difference between the two, the distance of the dotted lines from the horizontal reference line indicates how large the difference is, with the difference being larger as the dotted lines are further from the reference line.

Before reporting of F0, intensity, and formants, however, we present findings from the analysis of duration and temporal alignment over the files.

### 3.1. *Temporal aspects*

As mentioned, a temporal alignment difference was noted over the whole recording session files (average recording session duration $\approx 178.2$ s per session). However, most phonetic analyses only include smaller files of several seconds in duration (average utterance length

≈ 2.02 s per utterance). Therefore, we report the utterance-level duration findings first in 3.1.1, and the temporal alignment issue in the longer recording session files in 3.1.2.

### 3.1.1. Utterance duration

Impressionistically, temporal differences at the utterance level were not directly observable when conducting the manual segmentation stage, so the discrepancies were expected to be small overall for files of that duration. We note that it was slightly more difficult to annotate files produced by the two Zoom methods which could be attributable to, for example, less clear information in the spectrogram and waveform.

On average, as shown in Figure 3, utterances recorded by AVR were shorter than H6 by 2.1 ms; Zoom-default were shorter by 3.3 ms; and Zoom-raw utterances were shorter by 11.4 ms. This level of temporal difference is usually negligible in most phonetic studies since human annotation differences can sometimes be much larger.

We also ran a linear mixed-effect model to test whether these utterance-level temporal differences were statistically significant. Results showed no significant difference for any recording method, as presented in Table 3.

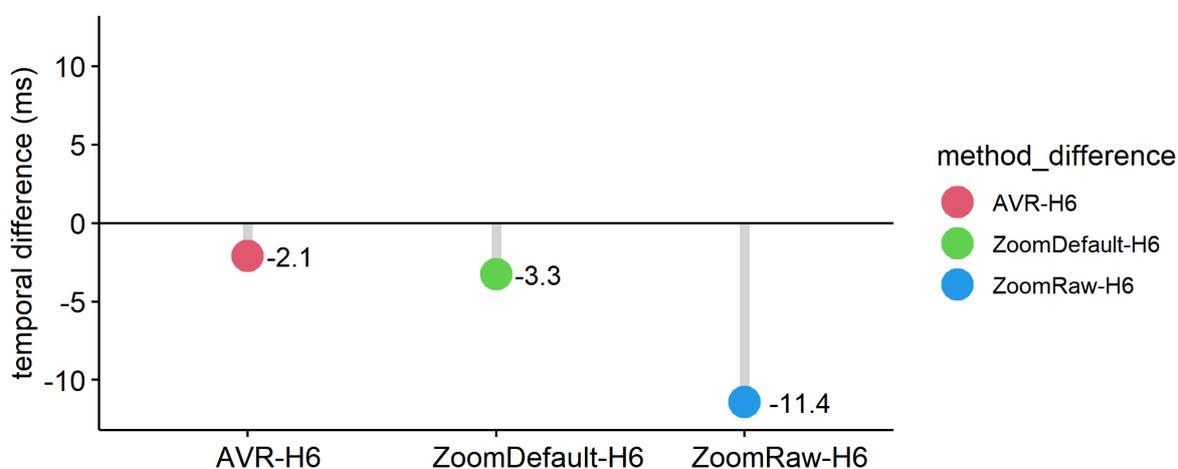

Figure 3: Mean utterance duration differences between AVR, Zoom-default, Zoom-raw and H6.

Table 3: Results from the linear mixed effect model for duration (intercept, H6). Final model: duration ~ method + (1|repetition) + (1|subject) + (1|utterance).

|  | Estimate | SE | df | t | Pr(> |t|) |
|---|---|---|---|---|---|
| (Intercept) | 2.018 | 0.177 | 7.5 | 11.424 | < 0.001 |
| methodAVR | 0.002 | 0.017 | 466.0 | 0.125 | 0.901 |
| methodZoom-default | -0.001 | 0.017 | 466.0 | -0.069 | 0.945 |
| methodZoom-raw | -0.009 | 0.017 | 466.0 | -0.552 | 0.581 |

### *3.1.2. Temporal alignment issue*

While the temporal difference is negligible on the utterance level, when a recording session becomes longer, the temporal difference between files becomes increasingly larger. To investigate the observed misalignment over time between recording methods, we chose three time points in each recording session file to compare across recordings, as reported in Section 2.5.1.

The temporal difference data, that is, the time point value of a comparison method (AVR/ Zoom-default/ Zoom-raw) minus the time point value of the baseline H6, are shown in Figure 4. The ms values reported are averages across speakers. AVR recordings started by having an earlier start than the H6 of 0.4 ms, and by the end, they were later than H6 recordings by 1.64 ms. Over the course of three minutes or more (i.e., the duration of our recording session files), this difference is extremely small for the AVR recordings, and we consider it to be trivial. Zoom-default, on the other hand, was earlier than H6 at the early time point by 3.95 ms. Over time, the temporal difference gradually became larger, and reached -27.55 ms by the last time point. Zoom-raw had a similar temporal difference and reached -25.65 ms by the last time point.

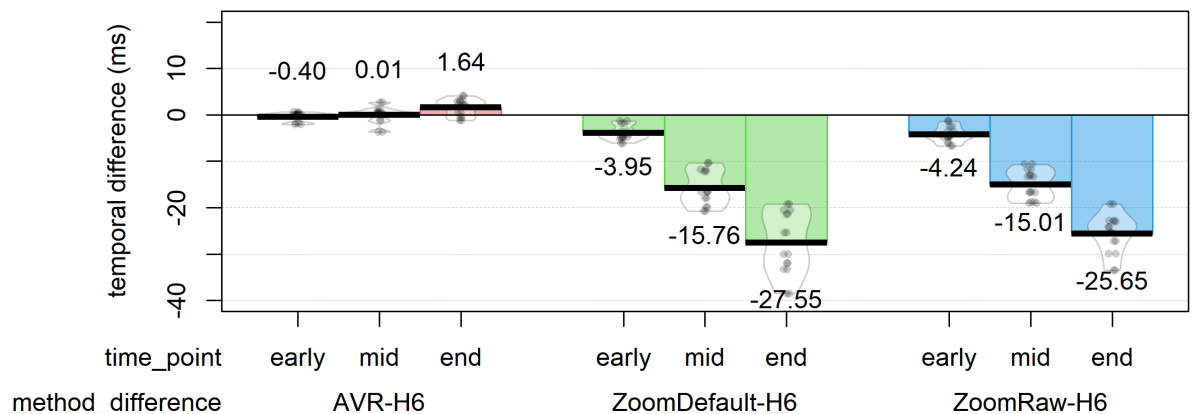

Figure 4: Temporal difference between AVR, Zoom-default, Zoom-raw and H6 at different time points.

The linearity between the temporal difference and time in H6 recording was tested with three separate simple linear models. The results of the temporal difference for AVR and the time of H6 suggest that they did form a linear relationship ($t = 4.49$, $p < 0.001$); however, the adjusted $R^2$ was only 0.289, which indicated that only 28.9% of the data were explained by the linear model. The result from the Zoom-default was $t = -37.21$, $p < 0.001$, adjusted $R^2 = 0.968$, which suggested a strong linear relationship between the Zoom-default temporal difference and the H6 time. Similarly, Zoom-raw reported $t = -45.66$, $p < 0.001$, adjusted $R^2 = 0.978$, also indicating a strong linear relationship. The linearity can be observed in Figure 5: as the time of H6 proceeds (i.e., at later points in an audio file), the temporal difference becomes larger in a linear fashion, especially for the two Zoom methods.

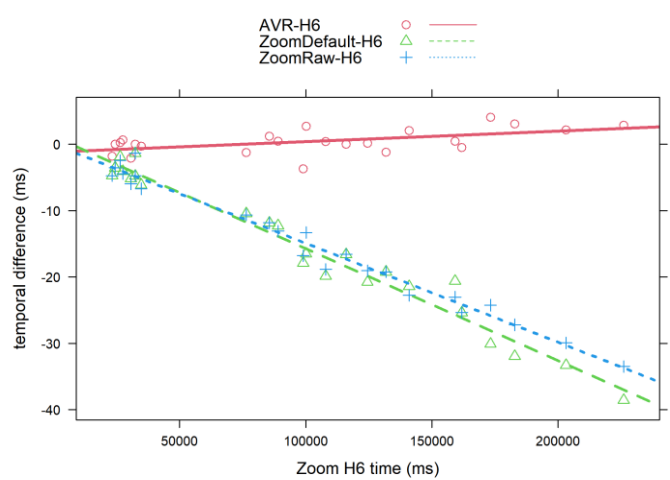

Figure 5: Linear regression of the temporal difference and Zoom H6 time.

### 3.2. Utterances

### 3.2.1. F0

Figure 6 shows the average F0 contours for each method across all utterances, speakers and repetitions. As indicated in Figure 6, the contours from each recording method are very similar overall. From the QGAMMs analysis, there is no evidence that utterance F0 contours from AVR, Zoom-default or Zoom-raw differed significantly from those of the H6. Firstly, there is no significant difference in the parametric coefficient (i.e., intercept height in Table 4). Further, for the smooth terms ("approximate significance of smooth terms" in Table 4), no difference was significant. These results are reflected in the difference plots (Figure 7) that show the confidence intervals for the AVR, Zoom-default and Zoom-raw difference curves are overlapped with the horizontal y = 0 line. Although towards the end of the curve, the two Zoom measures are not overlapping with the y = 0 line, indicating a small difference, the overall difference did not reach significance according to the results in Table 4.

In the following subsections, difference plots are presented for each measurement, and model predictions are shown for the ways in which the contours vary from the predicted contour for H6; where y = 0, the two contours do not differ in shape. In Figure 7, the first panel shows the model prediction for the H6 contour, taking into account speaker and utterance variability, followed by the difference plots for AVR, Zoom-default and Zoom-raw. In subsequent plots, only the difference curves are presented. Significant results are annotated in the plot.

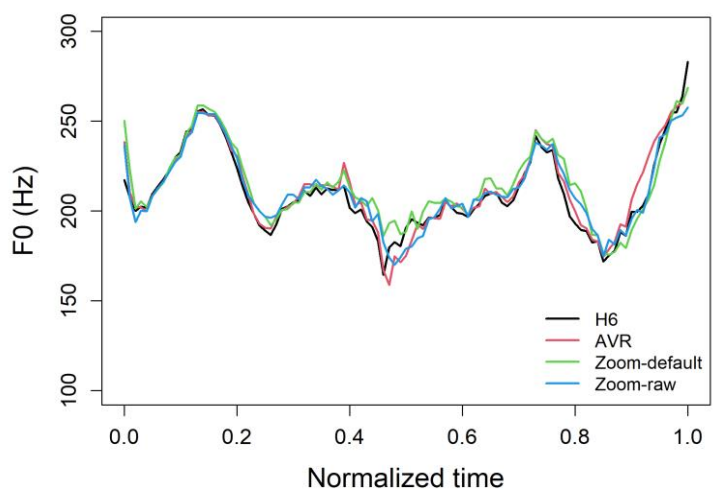

Figure 6: Average utterance F0 (Hz) contours by method, across all speakers, utterances and repetitions.

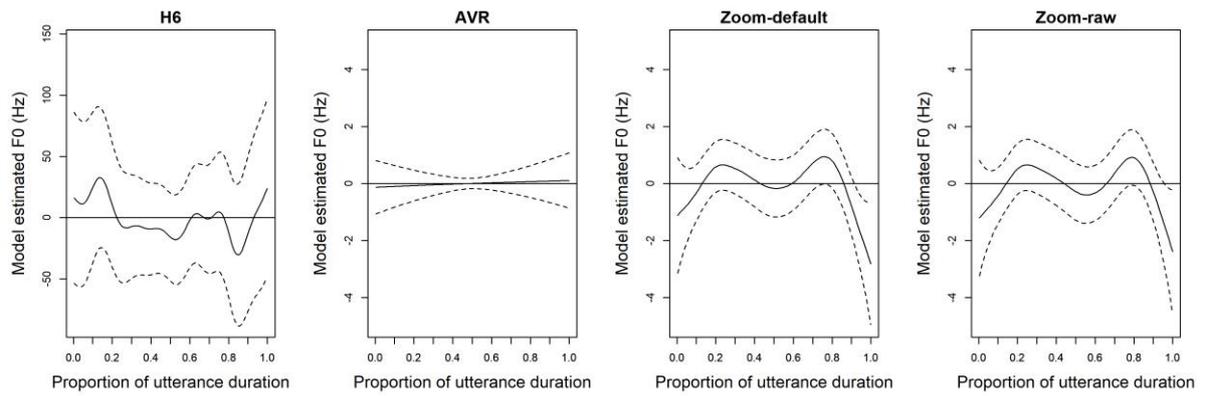

Figure 7: Predicted utterance F0 contour for H6 (left) and difference plots for utterance F0 for the three test methods (AVR difference curve center left; Zoom-default difference curve center right; Zoom-raw difference curve right).

Table 4: Summary table of utterance F0 QGAMM. Final model: f0 ~ gender + method + s(measurement.no, k = 20) + s(measurement.no, by = methodOrd, k = 20) + s(measurement.no, speaker, bs = "fs", m = 1, k = 20) + s(measurement.no, utterance_id, bs = "fs", m = 1, k = 20)

| Parametric coefficients: | | | | |
|---|---|---|---|---|
| | Estimate | Std. Error | z value | Pr($>$|z|) |
| (Intercept) | 266.16522 | 25.96188 | 10.252 | $<$2e-16 |
| genderM | -139.09013 | 9.72814 | -14.298 | $<$2e-16 |
| methodAVR | 0.06214 | 0.26893 | 0.231 | 0.817 |
| methodZoom-default | -0.05097 | 0.27186 | -0.187 | 0.851 |
| methodZoom-raw | -0.00886 | 0.27064 | -0.033 | 0.974 |
| Approximate significance of smooth terms: | | | | |
| | edf | Ref.df | Chi.sq | p-value |
| s(measurement.no) | 12.802 | 13.479 | 24.415 | 0.0260 |
| s(measurement.no):methodOrdAVR | 1.077 | 1.148 | 0.058 | 0.8973 |
| s(measurement.no):methodOrdZoom-default | 4.747 | 5.929 | 12.870 | 0.0431 |
| s(measurement.no):methodOrdZoom-raw | 4.880 | 6.093 | 10.951 | 0.0930 |
| s(measurement.no,speaker) | 137.133 | 158.000 | 26734.829 | $<$2e-16 |
| s(measurement.no,utterance_id) | 87.625 | 99.000 | 76212.693 | $<$2e-16 |

### 3.2.2. Intensity

As shown in Figure 8 (left panel), raw intensity values exhibited an overall height difference. However, the difference was expected since the different methods were placed at different distances from the speaker's mouth. H6 had overall higher intensity than the three comparison methods, and this was supported in the analysis of the data (see Supplementary Material 2).[4] Therefore, we normalized the intensity data (discussed in Section 2.5) and the analysis of those data is presented below. Despite normalization, contour intercept height was significantly higher for AVR (see Table 5). The contour shapes for all comparison methods were also significantly different from the H6, as seen in the significant results for the smooth terms (Table 5). From inspecting the difference plots (Figure 9), we can see that the differences are greatest at the beginning of the contour for all methods (where the curve is furthest from the y = 0 line). Zoom-default and Zoom-raw quickly rise to relatively consistent differences from H6, while AVR gradually rises over time. Visual inspection of the difference plots (Figure 9) and average contours (Figure 8, right panel) together suggests that the intensity differences are most different for the AVR at around measurement point 0.5 and 0.8, for Zoom-default at around 0.5, and for Zoom-raw, 0.8. These are points when the normalized intensity of the respective recording method is higher than that of H6.

---

[4] Also available through the project's OSF repository https://osf.io/34m5s/

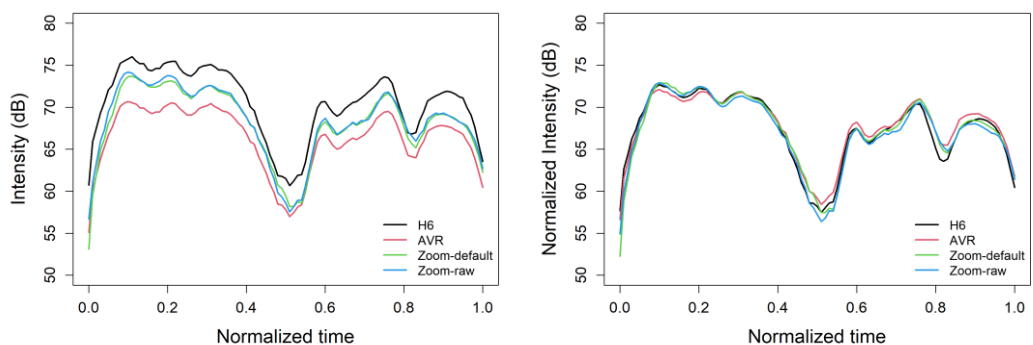

Figure 8: Average utterance intensity (dB, left) and normalized intensity (dB, right) contours by method, across all speakers, utterances and repetitions.

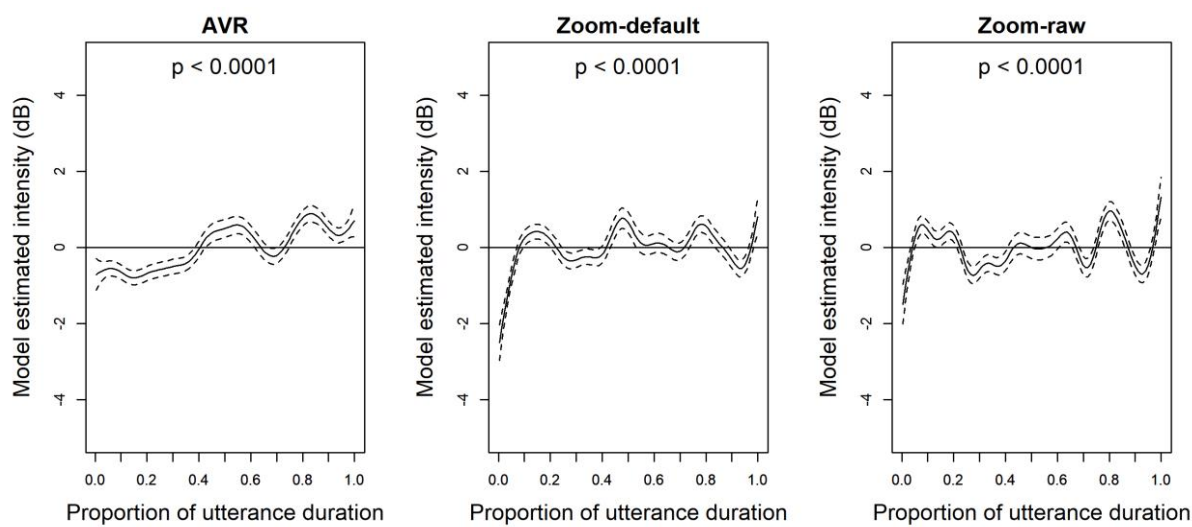

Figure 9: Difference plots for utterance normalized intensity (AVR difference curve left; Zoom-default difference curve center; Zoom-raw difference curve right).

Table 5: Summary table of normalized utterance intensity QGAMM. Final model: intensity ~ method + s(measurement.no, k = 20) + s(measurement.no, by = methodOrd, k = 20) + s(measurement.no, speaker, bs = "fs",  m = 1, k = 20) + s(measurement.no, utterance_id, bs = "fs", m = 1, k = 20)

| Parametric coefficients: | | | | |
|---|---|---|---|---|
| | Estimate | Std. Error | z value | Pr(> \|z\|) |
| (Intercept) | 78.09171 | 4.59853 | 16.982 | < 2e-16 |
| methodAVR | 0.22847 | 0.03138 | 7.281 | 3.32e-13 |
| methodZoom-default | 0.01382 | 0.03140 | 0.440 | 0.659869 |
| methodZoom-raw | -0.12679 | 0.03262 | -3.887 | 0.000102 |
| Approximate significance of smooth terms: | | | | |
| | edf | Ref.df | Chi.sq | p-value |
| s(measurement.no) | 16.84 | 17.21 | 1043.6 | <2e-16 |
| s(measurement.no):methodOrdAVR | 12.22 | 14.56 | 320.0 | <2e-16 |
| s(measurement.no):methodOrdZoom-default | 15.37 | 17.37 | 263.0 | <2e-16 |
| s(measurement.no):methodOrdZoom-raw | 16.80 | 18.30 | 255.7 | <2e-16 |
| s(measurement.no,speaker) | 135.06 | 159.00 | 8928.4 | <2e-16 |
| s(measurement.no,utterance_id) | 87.25 | 99.00 | 84438.2 | <2e-16 |

### 3.3. Vowels

#### 3.3.1. Formants

Here, F1 and F2 formant values are considered for all vowels together to investigate if there is an overall effect of recording method that results in higher or lower values.[5] Vowel spaces based on /iː, æ, uː/ for all participants, and by gender are considered in Section 3.3.2.

Regarding F1, there were significant differences in intercept as well as contour shape and overall contour height (Table 6). As can be seen in the average contour plots (Figure 10), the three comparison methods had overall lower F1 values than the H6. Note however, that there are differences over time, for example the Zoom-default contour had the lowest starting value, followed by a quick rise to approximately the H6 level, then declined at a slower rate than the H6, which resulted in a higher value at the end (see also the difference plots in Figure 11, discussed further, below). In terms of height, AVR has the smallest intercept difference; -11 ± 2 Hz ($p < 0.0001$). Zoom default had a difference of -17 ± 2 Hz ($p < 0.0001$), while

---

[5] Vowel F0 results are available in Supplementary Material 3.

Zoom-raw had the largest difference of -43 ± 2 Hz ($p < 0.0001$). The significant smooth terms for all comparison methods suggest that the F1 contours are different in shape from the H6 contours. These differences are reflected in the difference plots (Figure 11). The AVR difference curve is close to the y = 0 line, suggesting that the AVR contour is similar in shape to the H6, though still differs over time. The two Zoom difference plots show that F1 contours for both methods are different from the H6.

Regarding F2, only the Zoom-default contour is significantly different from the H6, and this is observed in both the parametric coefficient and smooth terms (Table 7). The Zoom-default intercept is an estimated 15 ± 3 Hz ($p < 0.0001$) higher than H6. The smooth term (Table 7) shows that there is little overlap between the Zoom-default and the H6 contour, and the difference smooth plot (Figure 12) shows that over the course of the vowel, the difference decreases and then increases. This is in part due to the Zoom-default F2 values not declining at the same rate at H6; this can be observed in the average F2 plot (Figure 10). Whereas, both AVR and Zoom-raw are similar in contour shape and height to the H6 in the average contour plot (Figure 10), and the difference plots (Figure 11). Note that for all methods, edf is near 1, suggesting a near linear relationship, as reflected in the relatively straight lines in the difference plots.

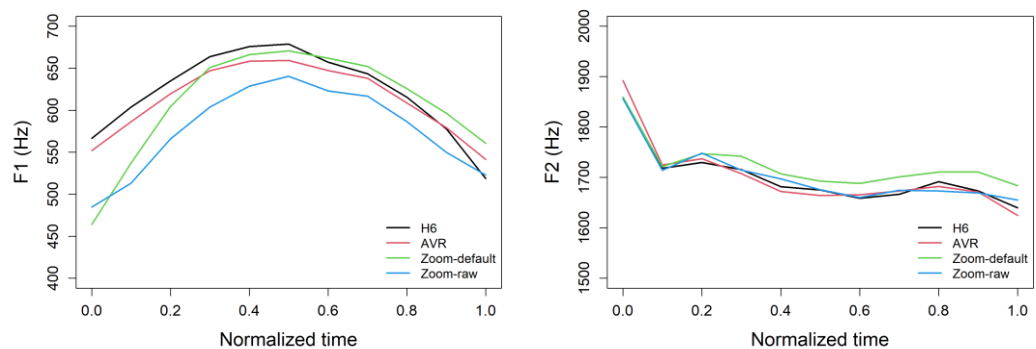

Figure 10: Average vowel F1 (left) and F2 (right) by method, across all speakers, vowels, and repetitions.

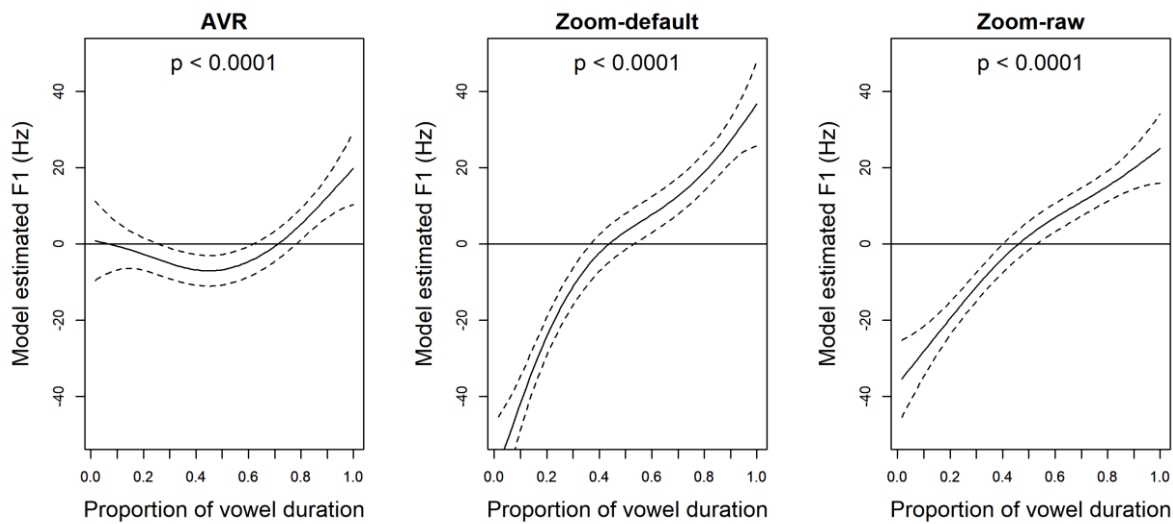

Figure 11: Difference plots for F1 over all vowels (AVR difference curve left; Zoom-default difference curve center; Zoom-raw difference curve right).

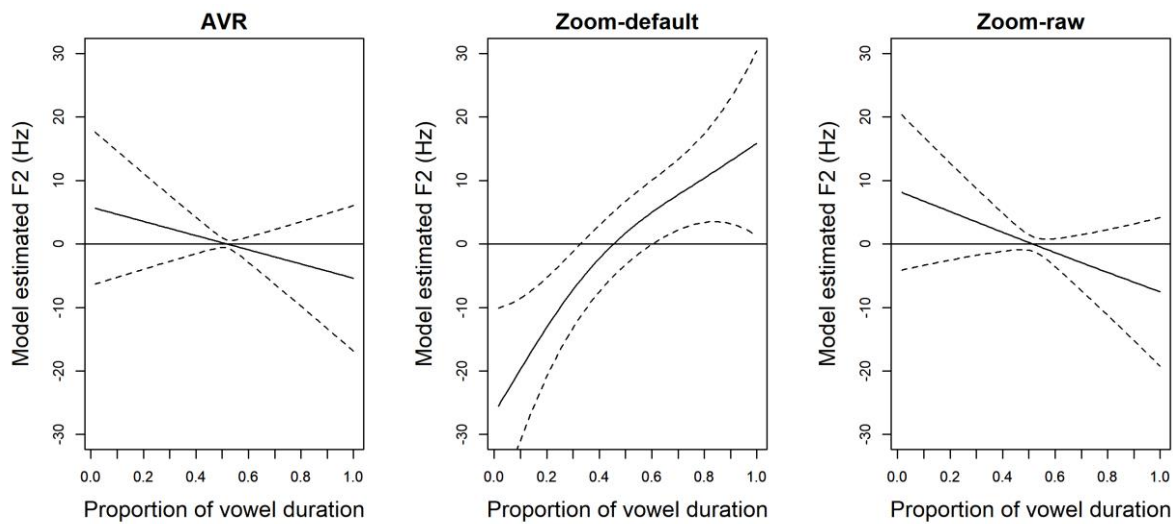

Figure 12: Difference plots for F2 over all vowels (AVR difference curve left; Zoom-default difference curve center; Zoom-raw difference curve right).

Table 6: Summary table of combined vowel F1 QGAMM. Final model: f1 ~ gender + method + s(measurement.no) + s(measurement.no, by = methodOrd) + s(measurement.no, speaker, bs = "fs", m = 1) +     s(measurement.no, vowel_id, bs = "fs", m = 1)

| Parametric coefficients: | | | | |
|---|---|---|---|---|
| | Estimate | Std. Error | z value | Pr($>$\|z\|) |
| (Intercept) | 682.777 | 23.562 | 28.977 | < 2e-16 |
| genderM | -155.259 | 16.446 | -9.441 | < 2e-16 |
| methodAVR | -11.054 | 1.516 | -7.290 | 3.09e-13 |
| methodZoom-default | -17.144 | 1.584 | -10.821 | < 2e-16 |
| methodZoom-raw | -42.937 | 1.638 | -26.220 | < 2e-16 |
| **Approximate significance of smooth terms:** | | | | |
| | edf | Ref.df | Chi.sq | p-value |
| s(measurement.no) | 7.009 | 7.466 | 160.16 | < 2e-16 |
| s(measurement.no):methodOrdAVR | 2.885 | 3.588 | 23.56 | 8.59e-05 |
| s(measurement.no):methodOrdZoom-default | 3.629 | 4.493 | 197.70 | < 2e-16 |
| s(measurement.no):methodOrdZoom-raw | 2.274 | 2.835 | 98.82 | < 2e-16 |
| s(measurement.no,speaker) | 48.519 | 70.000 | 1997.66 | < 2e-16 |
| s(measurement.no,vowel_id) | 178.676 | 233.000 | 47776.86 | < 2e-16 |

Table 7: Summary table of combined vowel F2 QGAMM. Final model: f2 ~ gender + method + s(measurement.no) + s(measurement.no, by = methodOrd) + s(measurement.no, speaker, bs = "fs", m = 1) + s(measurement.no, vowel_id, bs = "fs", m = 1)

| Parametric coefficients: | | | | |
|---|---|---|---|---|
| | Estimate | Std. Error | z value | Pr(>\|z\|) |
| (Intercept) | 1851.556 | 60.490 | 30.609 | < 2e-16 |
| genderM | -251.629 | 51.626 | -4.874 | 1.09e-06 |
| methodAVR | 1.909 | 3.307 | 0.577 | 0.564 |
| methodZoom-default | 15.125 | 3.250 | 4.654 | 3.26e-06 |
| methodZoom-raw | 5.230 | 3.317 | 1.577 | 0.115 |
| **Approximate significance of smooth terms:** | | | | |
| | edf | Ref.df | Chi.sq | p-value |
| s(measurement.no) | 5.833 | 6.342 | 20.796 | 0.00271 |
| s(measurement.no):methodOrdAVR | 1.009 | 1.018 | 0.894 | 0.34900 |
| s(measurement.no):methodOrdZoom-default | 1.735 | 2.160 | 12.931 | 0.00236 |
| s(measurement.no):methodOrdZoom-raw | 1.032 | 1.062 | 1.719 | 0.19354 |
| s(measurement.no,speaker) | 42.374 | 70.000 | 2694.238 | < 2e-16 |
| s(measurement.no,vowel_id) | 189.030 | 233.000 | 60533.855 | < 2e-16 |

### 3.3.2. *Vowel spaces*

To investigate if the methods altered the shape of the vowel space, that is, have different effects on formants depending on the Hz range in which they occur, we considered the three vowels /iː, æ, uː/. These vowels are representative of height and backness differences in the vowel space, and approximations of somewhat similar vowels were observed in the speech of all eight participants. Average contours are visualized in Figure 13, while the difference plots for F1 are presented in Figure 14, and for F2 in Figure 15. A successive analysis, below (also included in the vowel analysis in Supplementary Materials 2), investigated the vowel spaces for women and men separately.

Regarding F1, for /iː/, there was no significant difference in contour shape for any method nor a significant difference in contour height at the intercept for AVR. However, there were significant differences for the Zoom methods; the F1 intercept for Zoom-default was significantly lower than the H6 by an estimated -55 ± 8 Hz ($p = < 0.0001$), and was lower for Zoom-raw by an estimate -56 ± 8 Hz ($p < 0.0001$). For /æ/ F1, there were no significant differences in contour height at the intercept, but there was a significant shape difference for Zoom-default and Zoom-raw (see difference curves in Figure 14, mid panel, and average contours in Figure 13, center panel.) For /uː/ F1, there were no significant differences in contour shape, but there was a significant difference in contour height at the intercept for Zoom-default and Zoom-raw, which in both cases resulted in lower values than the H6; -36 ± 9 Hz $p < 0.0001$ for Zoom-default, and -43 ± 8 Hz, $p < 0.0001$. Interestingly, as can be observed in the difference plots (Figure 14, lower panel) and edf values (Table 8), differences between methods and H6 for F1 of the two high vowels /iː, uː/ is nearly linear, whereas differences are not linear for F1 of the low vowel /æ/ for the two Zoom methods. Further, for the two high vowels, both Zoom methods resulted in consistently lower values than the H6.

Regarding F2, for /iː/ there were no significant height differences (Table 8). The approximate significance of smooth terms, as well as difference plots (Figure 15, upper panel) suggest that both Zoom-default and Zoom-raw contours are significantly different from the H6, and for Zoom-default, this is nearly a linear relationship. F2 of /æ/ did not differ significantly in any respect for any method. For /uː/, there was a significant difference in contour intercept height for Zoom-default and Zoom-raw such that both methods resulted in higher F2 values than H6; Zoom-default had higher F2 by an estimated 79 ± 12 Hz ($p < 0.0001$), Zoom-raw had higher F2 by an estimated 59 ± 14 Hz ($p < 0.0001$). There were no significant contour shape differences.

Through visualizing these results in the F1-F2 space, it is clear that the formant differences for /iː, æ, uː/ result in different vowel spaces captured by the recording methods, and that the differences change over time. Figure 16 shows average values across speakers and repetitions for /iː, æ, uː/ at 10% (a), 50% (b), and 90% (c) through the vowel. Overall, we can see that AVR patterns with H6, and these differ from the two Zoom methods. Further, the two Zoom methods provide lower values for the F1 of the high vowels /iː/ and /uː/, which is evidenced by the lower intercept (Table 8). The three plots from different time points in the vowels' productions show that over the time course, the vowel space of the two Zoom methods, as compared to that of H6, is first compressed and then expanded. There are two predominant sources of this: first, /æ/'s F1 of the two Zoom methods is first lower than that of H6, and then gradually becomes higher (Figure 14, mid panel). Second, /iː/'s F2 from the two Zoom methods is also first lower than that of H6, and then gradually becomes higher (Figure 15, upper panel).

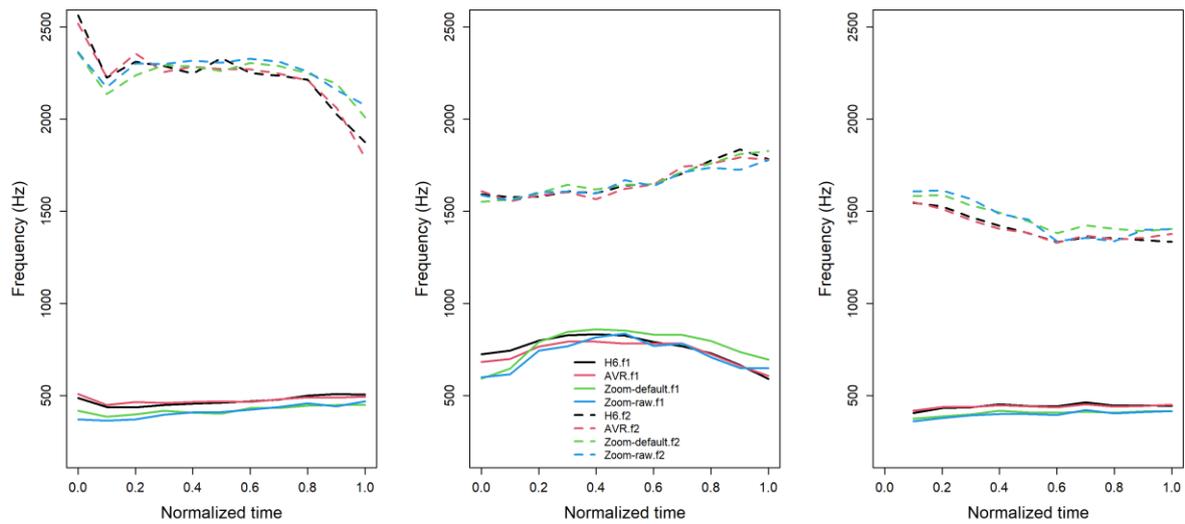

Figure 13: Average contours for F1 and F2 (Hz) for /iː/ (left), /æ/ (center) and /uː/ (right) vowels, by method, across all speakers and repetitions.

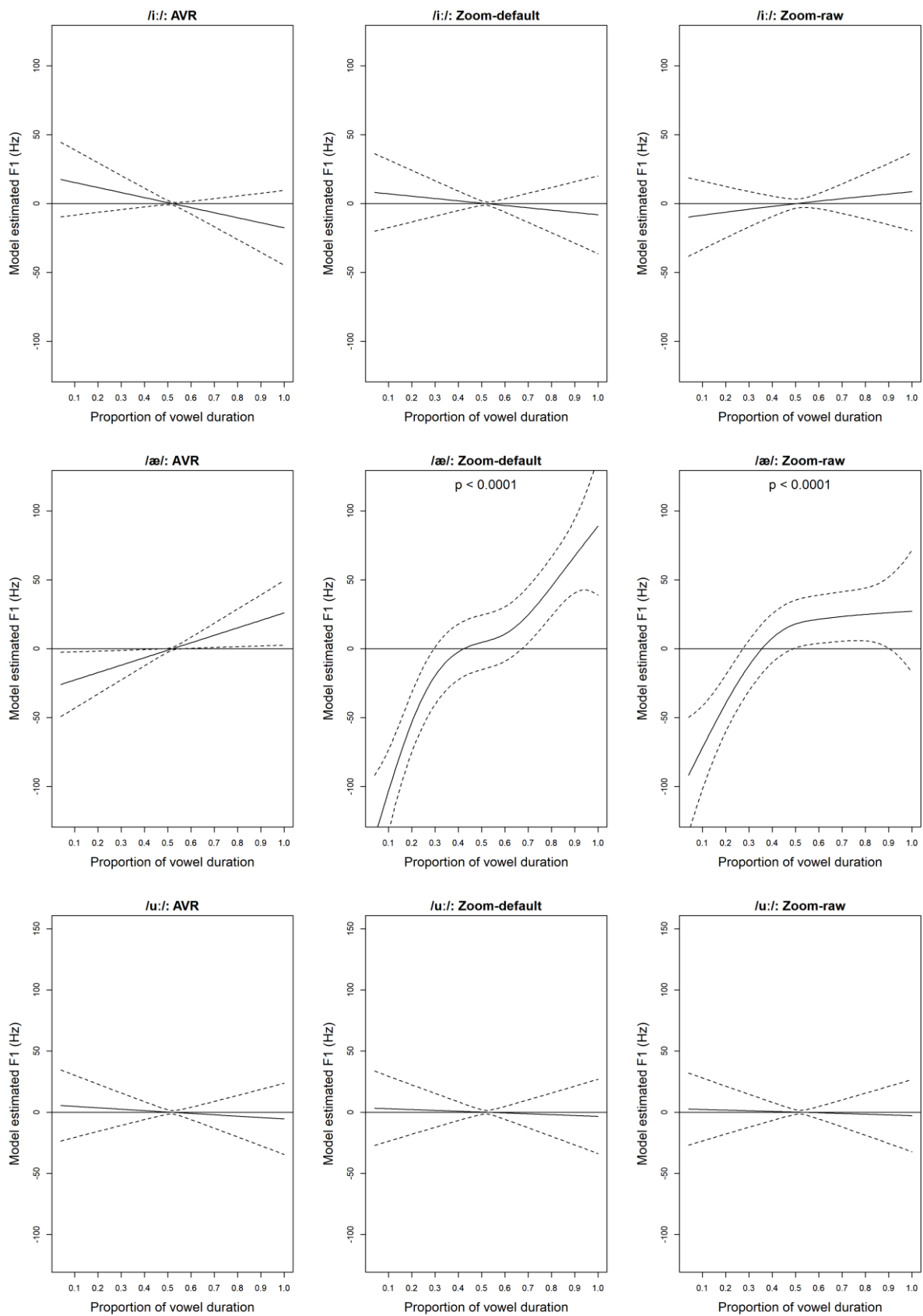

Figure 14: Difference plots for F1 for /iː/ (top), /æ/ (middle) and /uː/ (bottom); AVR left; Zoom-default center; Zoom-raw right.

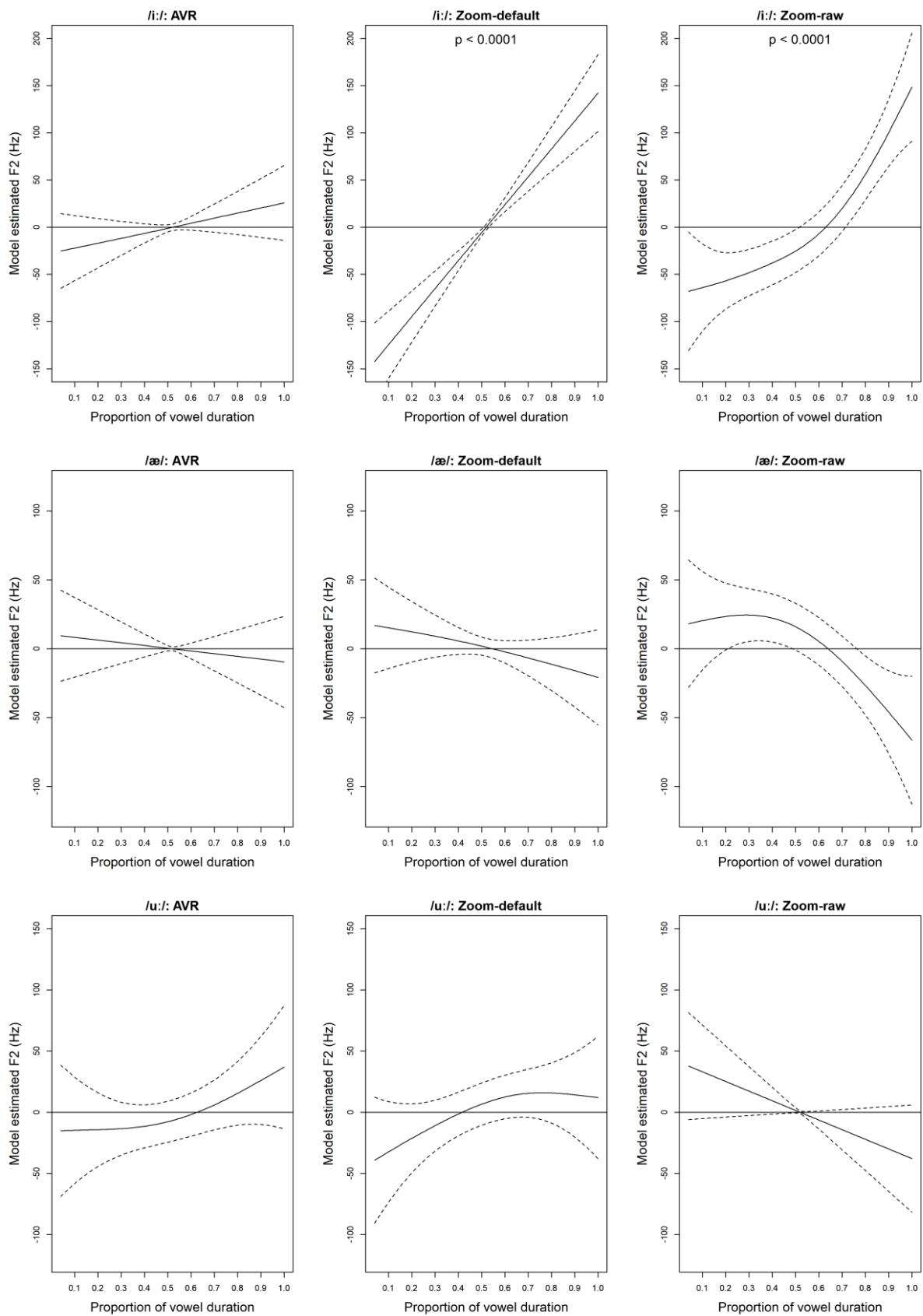

Figure 15: Difference plots for F2 for /iː/ (top), /æ/ (middle) and /uː/ (bottom); AVR left; Zoom-default center; Zoom-raw right.

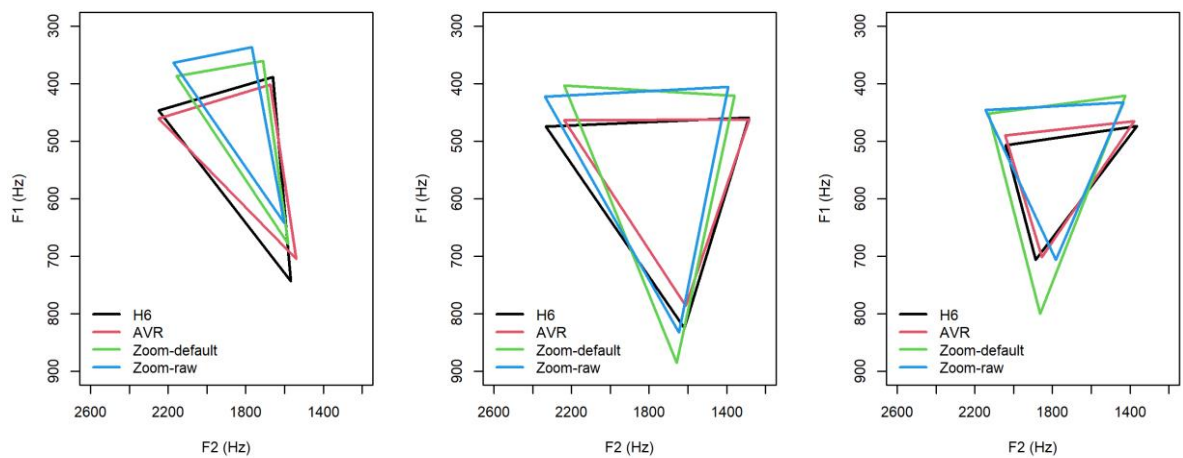

Figure 16: First and second formant values (Hz) for /i:/, /æ/ and /u:/ at the 0.1 time point (left), midpoint (center) and 0.9 (right) time point, plotted in F1-F2 vowel space.

Table 8: Summary table of /i:, æ, u:/ F1 and F2 QGAMM. Final model: formantValue ~ gender + formantVow + s(measurement.no, by = formantVow) + s(measurement.no, by = IsAVR.f1.a) + IsAVR.f1.a + s(measurement.no, by = IsAVR.f1.i) + IsAVR.f1.i + s(measurement.no, by = IsAVR.f1.u) + IsAVR.f1.u + s(measurement.no, by = IsAVR.f2.a) + IsAVR.f2.a + s(measurement.no, by = IsAVR.f2.i) + IsAVR.f2.i + s(measurement.no, by = IsAVR.f2.u) + IsAVR.f2.u + s(measurement.no, by = IsZoom.default.f1.a) + IsZoom.default.f1.a + s(measurement.no, by = IsZoom.default.f1.i) + IsZoom.default.f1.i + s(measurement.no, by = IsZoom.default.f1.u) + IsZoom.default.f1.u + s(measurement.no, by = IsZoom.default.f2.a) + IsZoom.default.f2.a + s(measurement.no, by = IsZoom.default.f2.i) + IsZoom.default.f2.i + s(measurement.no, by = IsZoom.default.f2.u) + IsZoom.default.f2.u + s(measurement.no, by = IsZoom.raw.f1.a) + IsZoom.raw.f1.a + s(measurement.no, by = IsZoom.raw.f1.i) + IsZoom.raw.f1.i + s(measurement.no, by = IsZoom.raw.f1.u) + IsZoom.raw.f1.u + s(measurement.no, by = IsZoom.raw.f2.a) + IsZoom.raw.f2.a + s(measurement.no, by = IsZoom.raw.f2.i) + IsZoom.raw.f2.i + s(measurement.no, by = IsZoom.raw.f2.u) + IsZoom.raw.f2.u + s(measurement.no, speaker, bs = "fs", m = 1) + s(measurement.no, word2, bs = "fs", m = 1)

| Parametric coefficients: | | | | |
|---|---|---|---|---|
| | Estimate | Std. Error | z value | Pr(> \|z\|) |
| (Intercept) | 903.151 | 31.333 | 28.825 | < 2e-16 |
| genderM | -202.627 | 28.674 | -7.067 | 1.59e-12 |
| formantVowf1.i | -352.403 | 38.660 | -9.115 | < 2e-16 |
| formantVowf1.u | -375.441 | 8.693 | -43.188 | < 2e-16 |
| formantVowf2.æ | 880.827 | 8.163 | 107.905 | < 2e-16 |
| formantVowf2.i | 1431.460 | 39.069 | 36.639 | < 2e-16 |
| formantVowf2.u | 611.806 | 10.978 | 55.729 | < 2e-16 |
| IsAVR.f1.a1 | -21.248 | 6.665 | -3.188 | 0.001433 |
| IsAVR.f1.i1 | 10.669 | 7.852 | 1.359 | 0.174219 |
| IsAVR.f1.u1 | -2.913 | 8.246 | -0.353 | 0.723907 |
| IsAVR.f2.æ1 | 4.144 | 9.306 | 0.445 | 0.656085 |
| IsAVR.f2.i1 | 3.596 | 10.886 | 0.330 | 0.741116 |
| IsAVR.f2.u1 | 0.792 | 13.107 | 0.060 | 0.951816 |
| IsZoom.default.f1.æ1 | 11.572 | 7.368 | 1.570 | 0.116303 |
| IsZoom.default.f1.i1 | -55.329 | 8.289 | -6.675 | 2.47e-11 |
| IsZoom.default.f1.u1 | -35.947 | 8.664 | -4.149 | 3.34e-05 |
| IsZoom.default.f2.æ1 | -3.275 | 9.377 | -0.349 | 0.726903 |
| IsZoom.default.f2.i1 | 3.698 | 11.107 | 0.333 | 0.739161 |
| IsZoom.default.f2.u1 | 78.956 | 11.963 | 6.600 | 4.12e-11 |
| IsZoom.raw.f1.æ1 | -25.915 | 7.523 | -3.445 | 0.000571 |
| IsZoom.raw.f1.i1 | -55.882 | 8.095 | -6.903 | 5.09e-12 |
| IsZoom.raw.f1.u1 | -43.226 | 8.473 | -5.102 | 3.37e-07 |
| IsZoom.raw.f2.æ1 | 5.805 | 9.425 | 0.616 | 0.537955 |
| IsZoom.raw.f2.i1 | 18.377 | 11.171 | 1.645 | 0.099975 |
| IsZoom.raw.f2.u1 | 59.192 | 13.507 | 4.382 | 1.17e-05 |
| **Approximate significance of smooth terms:** | | | | |
| | **edf** | **Ref.df** | **Chi.sq** | **p-value** |
| s(measurement.no):formantVowf1.æ | 5.752 | 6.778 | 195.124 | < 2e-16 |
| s(measurement.no):formantVowf1.i | 1.011 | 1.016 | 6.520 | 0.0109 |
| s(measurement.no):formantVowf1.u | 2.305 | 2.727 | 11.662 | 0.0119 |
| s(measurement.no):formantVowf2.æ | 4.148 | 5.014 | 70.934 | < 2e-16 |
| s(measurement.no):formantVowf2.i | 6.631 | 7.710 | 449.500 | < 2e-16 |
| s(measurement.no):formantVowf2.u | 4.421 | 5.365 | 100.002 | < 2e-16 |
| s(measurement.no):IsAVR.f1.æ1 | 1.005 | 1.009 | 4.962 | 0.0264 |
| s(measurement.no):IsAVR.f1.i1 | 1.006 | 1.012 | 1.687 | 0.1968 |
| s(measurement.no):IsAVR.f1.u1 | 1.009 | 1.018 | 0.138 | 0.7169 |
| s(measurement.no):IsAVR.f2.æ1 | 1.006 | 1.011 | 0.331 | 0.5698 |
| s(measurement.no):IsAVR.f2.i1 | 1.025 | 1.050 | 1.739 | 0.2010 |
| s(measurement.no):IsAVR.f2.u1 | 1.462 | 1.779 | 1.516 | 0.3127 |
| s(measurement.no):IsZoom.default.f1.æ1 | 3.333 | 4.134 | 56.622 | < 2e-16 |
| s(measurement.no):IsZoom.default.f1.i1 | 1.007 | 1.013 | 0.332 | 0.5690 |
| s(measurement.no):IsZoom.default.f1.u1 | 1.008 | 1.016 | 0.047 | 0.8421 |
| s(measurement.no):IsZoom.default.f2.æ1 | 1.130 | 1.243 | 1.636 | 0.3000 |
| s(measurement.no):IsZoom.default.f2.i1 | 1.021 | 1.042 | 49.325 | < 2e-16 |
| s(measurement.no):IsZoom.default.f2.u1 | 1.616 | 1.994 | 2.566 | 0.2586 |
| s(measurement.no):IsZoom.raw.f1.æ1 | 2.617 | 3.257 | 23.084 | 6.83e-05 |
| s(measurement.no):IsZoom.raw.f1.i1 | 1.049 | 1.093 | 0.374 | 0.5495 |
| s(measurement.no):IsZoom.raw.f1.u1 | 1.010 | 1.019 | 0.036 | 0.8808 |
| s(measurement.no):IsZoom.raw.f2.æ1 | 2.063 | 2.574 | 10.143 | 0.0135 |
| s(measurement.no):IsZoom.raw.f2.i1 | 2.421 | 3.017 | 32.645 | < 2e-16 |
| s(measurement.no):IsZoom.raw.f2.u1 | 1.002 | 1.004 | 2.993 | 0.0839 |
| s(measurement.no,speaker) | 27.335 | 70.000 | 643.185 | < 2e-16 |
| s(measurement.no,word2) | 28.461 | 70.000 | 1755.168 | < 2e-16 |

Further examination of the F1 and F2 data was performed, investigating female and male speech separately (see Figure 17 for average vowel spaces; female, left, and male, right; full result table available in Supplementary Material 2). The analyses showed that time-varying effects (i.e., contour shape differences) on F1 and F2 generally hold for both genders. However, there are intercept differences which patterned differently by gender. For example, for F1, the intercepts for /iː/ and /uː/ in the male data as recorded by both Zoom methods were not significantly different from the H6, whereas these were for the data overall, presumably because of the effect in the female data (/iː/ Zoom-default -69 ± 8 Hz, $p < 0.0001$; /iː/ Zoom-raw -73 ± 8 Hz, $p < 0.0001$; /uː/ Zoom-default -71 ± 13 Hz, $p < 0.0001$; /uː/ Zoom-raw -63 ± 13 Hz, $p < 0.0001$). The intercept for /æ/ F1 Zoom-raw data was significantly different from H6 in the male data, which is not found overall. For F2, a main difference was that the Zoom-raw intercept for /uː/ was not found to be significantly different from the H6 in the female data, but was in the male data (111 ± 13 Hz, $p < 0.0001$). Other intercept results conform to the patterns observed overall.

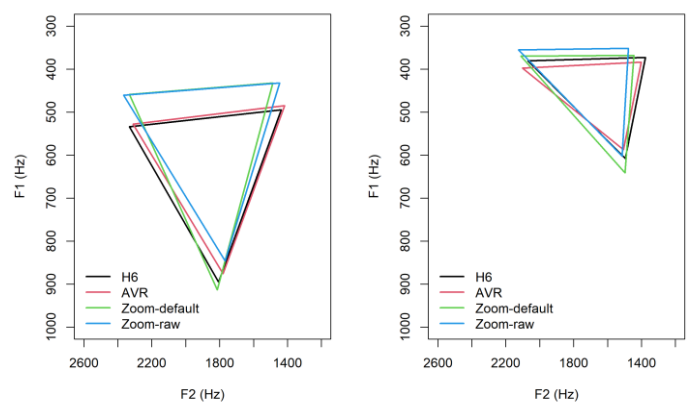

Figure 17: Average first and second formant values (Hz) for /iː/, /æ/ and /uː/, plotted in the F1-F2 vowel space for female data on the left, and male data on the right.

## 4. Discussion

In this study, we examined the effects of three remote recording methods on speech data, comparing them with simultaneous lab-quality baseline recordings, to assess their suitability in remote data collection for phonetic studies using speech data collected from eight participants. We evaluated AVR, a non-lossy format file recording smartphone application, and the conferencing app Zoom with and without post-processing, all used to record locally. Taking a dynamic approach, F0 and intensity contours of utterances, and vowel measures of F1, and F2 were modelled using QGAMMs to investigate the effects of recording method on each measure over the course of an utterance or vowel, respectively. Linear mixed effects methods were used to analyze temporal measures. In the following sections we review the findings, situating them in the literature, and venture some speculations as to the causes for what we observed.

### 4.1. Temporal measures

We examined both the duration of utterances across the recording methods and time points throughout the whole recording session files. While we found small differences in utterance duration between the comparison recording methods and the H6, these differences are considered to be negligible as they were in the range of 2 ms to 11 ms. Ladefoged (2003, p. 140) argued that duration measurements should only be reported to the nearest 5 ms since it is impossible to "make a reliable measurement of a duration in tenths of a millisecond". This suggests that the differences of 2.1 ms (AVR) and 3.3 ms (Zoom-default) from H6 are within an acceptable error range. Although Zoom-raw had the biggest difference in duration, it was only 11.4 ms on average. The differences in all three methods could be attributable to manual segmentation issues at the utterance boundaries; the values we observe are within the region of what has been found in inter-rater reliability studies (Machač & Skarnitzl, 2009, pp. 13–14). In any case, the statistical analysis did not reach significance. Therefore, we conclude that the duration captured by all methods did not pose an issue in analyzing the duration utterances. Furthermore, we suggest that the durational differences for individual words and segments would not be affected substantially using these three methods.

However, we did observe a timing issue in Zoom-default and Zoom-raw recordings over the long, recording session files. AVR files did not present a substantial temporal difference from H6 files overall, while files created using the two Zoom methods diverged from the H6 files in a linear fashion, with the difference increasing over time such temporal landmarks in the files created using the two Zoom methods were earlier than those in the H6 recordings, and became increasingly earlier over time. This temporal difference may affect the extraction of other measures when an audio data file is long. For example, the temporal alignment difference was revealed in our case when we attempted to use only one annotated TextGrid (from the H6) for each participant across all recording session files. The findings discussed above for the utterance files lead us to believe that if words or segments are investigated, the temporal difference would small enough to be negligible across the recording methods. We were confident in this assessment, and so used only one set of annotated files to extract F0 and formant measure for analysis (see Section 2.5.1 for details).

It is unclear what factors contributed to this discrepancy. Sanker et al. (2021) have also reported a similar issue in Zoom, Cleanfeed, and Messenger. However, from the figures they presented, the temporal difference did not seem to increase linearly through time, and was appreciably larger for Cleanfeed and Messenger than Zoom, which appeared negligible (as it was in the present study). Our speculation is that the temporal difference was caused by the compression and decompression of the files; because the difference is linear, we assume that something is affecting the audio in a small way consistently throughout a recording when using Zoom with any settings. The differences in the recording session files could also come about from how silences are treated by a Zoom algorithm, which was not made publicly available in any documentation on their website. Beyond unknown algorithms used by Zoom, in online discussions, Zoom users have posed questions about large differences between audio files recorded through Zoom, with suggested causes being that the mute button acts as a pause button in effect in recordings made in Zoom meetings, with participants only being recorded when not on mute. This does not appear to be the case in our data as the participants did not use the mute button, and we believe this would result in inconsistent differences, not the small linear increase in the difference we observe.

### 4.2. F0

In line with previous research, F0 over the utterances was found to be accurately captured by all recording methods. This finding suggests that prosody researchers can reliably use F0 recorded by AVR and Zoom with either setting. This applies to both static F0 measures (e.g., single F0 point or mean F0 values), as has been found in previous research (e.g. Fahed et al., 2022; Jannetts et al., 2019; Maryn et al., 2017; Vogel et al., 2015; Zhang et al., 2021) and when investigating F0 contour shape as we did here. While not significant, it was interesting to note that F0 contours from both Zoom methods had similar difference curves, suggesting that this is an effect of the Zoom software, and not participants' devices or due to individuals.

### 4.3. Intensity

Intensity contours showed considerable contour shape differences between the test methods and H6. Normalization only served to account for a height difference that reflected the distance the speakers were from the recording devices. The retention of the contour shape differences suggests that intensity values cannot simply be corrected for by normalization, as differences were not consistent over time (cf. Penney et al. 2021). For example, intensity as recorded by AVR increased over time, reflected in an average intensity contour that started lower than the H6 and concluded higher (Figure 8, right panel). The differences in shape for the test methods may pose an issue for prosody researchers who want to compare intensity between syllables, for example; it appears that using any of these three methods, the relative difference may not be accurately captured. Moreover, as pointed out by Zhang et al. (2021), Zoom-default presented periods of extremely reduced intensity occurring at random; an issue that persisted in the recordings analyzed in this paper. We speculate that this issue arises from a Zoom feature that is designed to remove background noise, inadvertently being applied to speech.

### 4.4. Formants

As was anticipated, formants were not always reliably tracked by the comparison methods. The combined vowel analysis showed F1 was overall lower for all test methods. The analysis of the three vowels /iː/, /æ/ and /uː/ revealed that the two Zoom methods resulted in lower intercept values for the two high vowels /iː/ and /uː/, and these values remained lower over time. In Zhang et al. (2021), the mean F1 value in Zoom (default mode) recordings was reported to be significantly lower for the eight cardinal vowels combined; in particular, the low vowels /a/ and /ɑ/ were observed to have the largest variance. While the results from Zhang et al. (2021) and the current study did not concur wholly, they both showed that the F1 values in Zoom recordings (with default or altered setting) were unreliable. The current dynamic analysis method provides a better understanding of the difference than what can be inferred from Zhang et al. (2021) and other studies considering static measures.

For F2 in the combined vowel analysis, only Zoom-default differed significantly from the H6 in the current study; Zoom-raw's difference from H6 did not reach statistical significance (p = 0.115), but was also worth noting. This result reflects the findings from Zhang et al. (2021), in which the mean F2 values were reported to have a significant difference for Zoom on default mode, and had the largest variance in the front vowels /i/, /e/, and /ɛ/. In the current study, /iː/ was found to have a contour difference in both Zoom-default and Zoom-raw, and /uː/ was different for Zoom-default and Zoom-raw such that both had higher F2 values than the H6 baseline recordings. These results again help us understand how F2 is captured by different recording methods in general, and alert us that even though Zoom-raw did not report any statistical difference in the combined vowel analysis, effects are observed for individual vowels, which in turn effects the shape of the vowel space overall.

Female and male speech was also found to be affected in different ways from each other, with the female vowel space overall being affected to a greater degree by recording method. F1 values of the high vowels were lower for females in data from the two Zoom methods, but not for males, while F1 of /æ/ was affected for males (significantly lower) in the Zoom-raw data but not females. F2 of /uː/ was not found to be affected by the two Zoom-raw in female data, but was significantly higher for males. The low accuracy of F1 seems to be related to the unknown Zoom algorithms. For instance, in Figure 18, the F2 curves in all three comparison methods are much closer to what is shown in the H6 panel than the F1 curves. In Zoom-raw, F1 almost disappears from the middle of the vowel, becoming increasingly less clear towards the end, but it is slightly better than Zoom-default. This may be related to Zoom's noise-cancellation algorithm in the default setting. Although Zoom-raw keeps the "original sound" as what the option suggests, as a conferencing application, it may still process the audio with complex algorithms which affects F1 more than F2.

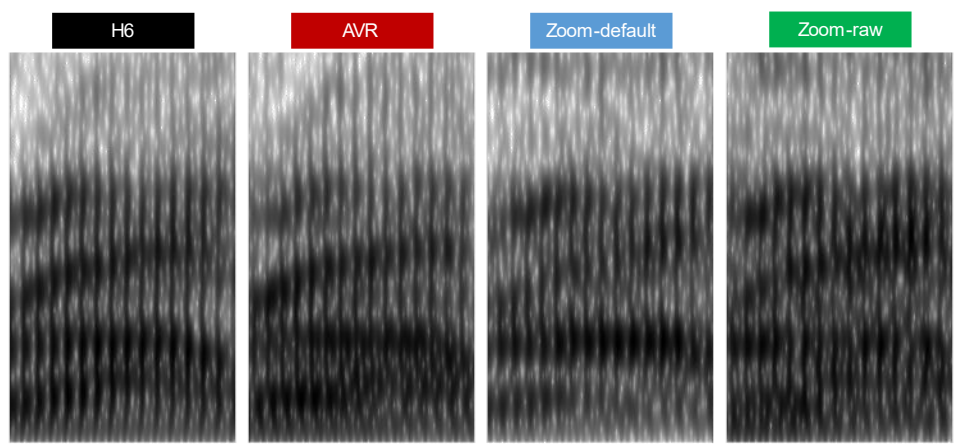

Figure 18: Spectrogram for the /ɑ/ vowel in 'ramen' produced by PM1.

Gender-based differences have been reported in other studies, such as Freeman & De Decker (2021b), with issues for the female speaker in the range of 750-1500 Hz. We did not find that frequencies in a particular range were affected (e.g., F2 of /uː/ was affected for males, but not for females, which show similar values in our data; see Figure 17), but our findings do suggest that female and male vowels are affected differently by recording method, especially online conferencing software options, as was also reported, though not statistically tested in Zhang et al. (2021) .

Overall, the results suggest that while the vowel space shape persists in data from all recording methods (cf. Freeman and De Decker's (2021a, 2021b), there are significant differences in individual vowels, particularly by the two Zoom methods. Crucially, the difference observed does not remain consistent over time. In contrast to Freeman and De Decker who found greatest differences for their female participant's low vowels, this study found that the high vowels were most affected. Our results concur in essence with Penney et al. (2021) who showed that for voice quality measures, method differences cannot be corrected for by procedures that linearly normalize extracted values, as the effect of method is different according to speaker gender and the value of the measurement of interest.

## 5. Conclusions and Recommendations

Through taking a dynamic approach to analysis, this study provides insight into the effects of recording methods on speech data for acoustic analysis that could not be offered by single time point or mean value analyses. The analyses of contour shape and height make it clear that different formant values from the comparison methods are not necessarily due to a consistent difference in tracking over the course of a segment, but reflect inaccuracies that differ over time. Because of this, we do not recommend the use of Zoom audio (with default, or original sound options) to investigate formants. AVR on the other hand, was seen to be more accurate in the values recorded for the vowels /iː, æ, uː/, and may be a suitable option for researchers as it is overall more similar to the H6 baseline recordings in contour height and shape. The comparison methods did not record utterance intensity accurately, even when overall height differences were accounted for by normalizing the values. Therefore, we do not recommend the use of these methods for the analysis of intensity and we believe this may also affect amplitude-based measures, discussed by Penney et al. (2021). AVR, Zoom-default, Zoom-raw are all suitable for tracking F0 with the analysis of utterances presented in this paper showing consistently tracked F0 across recording methods (see Supplementary Material 3 for the analysis of vowels).

With respect to the matched time points analysis, AVR was found to be the most consistently aligned with the H6 over the course of the full recordings, with the two Zoom methods having shorter duration between clap aligned points over the file. Issues, for example, are observed at the first time point, at approximately 25 seconds, and continue linearly over the course of recordings. However, for the annotation of small files (utterance size, for example, as used in this study) durational effects were not observed.

We recommend that, of the methods we tested, researchers record wav files through AVR on smartphones as a primary recording method. It may be that this method more generally reflects non-lossy recording options on smartphones; we make no claims about other software options, but we note their relative success in other studies (e.g. Penney et al., 2022), and the accessibility of smartphones to potential participant groups. In this study, we conducted a targeted evaluation of the AVR app, as it is accessible on both Android and iOS platforms. Our aim was to provide a single, considered recommendation regarding this application (and Zoom options) to researchers seeking a straightforward solution. Furthermore, we do not dismiss the validity of using other computer-based recording options, especially those that do not present the issues observed with online conferencing software. We do recommend that recording sessions are monitored using an online conferencing software such as Zoom, Skype, or Teams (Leemann et al., 2020). If backup recordings are made using Zoom, it is important to note that when using the Zoom-raw settings analyzed here, all speakers must wear headphones to avoid feedback when in a meeting with other people speaking. For file transferring, there are a range of free transfer platforms including WeTransfer and Send Anywhere. This process can be undertaken at the conclusion of a monitored recording session so that the experimenter can confirm their receipt.

This study has given further support for previous research, such that F0 is generally a reliable measure with the dynamics of its movement faithfully captured by the methods we tested and that formants pose difficulties. Further, these difficulties are different for different vowels and speakers. With respect to differences, this study has also provided more detail on how speech measures are affected by recording method, showing how, depending on the measure examined, single time point measures could be non-significantly different (e.g., F1 of /æ/, or F2 of /i/ taken at the midpoint, which are shown in the QGAMMs to overlap with the H6 baseline measure at this point), but that finding would represent a coincidence, rather than reliability of recording method. We draw attention to differences being often nonlinear across the measures we examined, and because of that, suggest that when performing analyses that compare measures compared at points across a word or utterance, especially intensity, that the results are best interpreted with caution.

We present our recommendations graphically in Figure 19. With all factors considered, AVR outperformed the two Zoom methods and was the most comparable recording method to the baseline H6 in our study. We therefore recommend this smartphone application to phonetic researchers for remote data collection, as part of a workflow that could include a Zoom meeting to facilitate tasks as well as help and monitor participants.

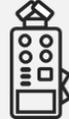

Figure 19: Recommendations for phonetics research based on results from the current study. Tick indicates that we recommend using the method; cross indicates that we do not recommend this method; tick with exclamation mark indicates that we consider this method being potentially suitable for some studies, but we advise researchers to consider the impact of the discrepancies on their specific studies and use with caution.

### Acknowledgements

This research was conducted with support from Grant No. ERC-ADG-835263, titled "Speech Prosody in Interaction: The form and function of intonation in human communication", awarded to Amalia Arvaniti. We are grateful for her support in planning and running the study, recording the demonstration audios, and her helpful advice in revising an early manuscript. We also thank Katherine Marcoux for her contribution to the project. We thank the editors (Katerina Iliopoulou, Yu-Jung Lin, and Lisa Davidson) and the two anonymous reviewers for their very helpful suggestions. For the purpose of Open Access, the authors have applied a CC BY public copyright license to any Author Accepted Manuscript (AAM) version arising from this submission.

**Appendix 1**

Summary of studies investigating the use of alternative recording methods for acoustic analysis (shaded cells were adapted based on Jannetts et al., 2019)

| Article | Number of speakers | Speaking task | Recording set-up | Devices used | Software | File formats | Acoustic measures |
|---------|-------------------|---------------|------------------|--------------|----------|-------------|-------------------|
| Vogel et al. (2015) | 15 speakers | Sustained vowel and reading passage | Simultaneous recordings with all devices | HDD plus table microphone; landline telephone; iPhone; computer plus headmounted microphone | | | F0 (mean, minimum, max, SD); Jitter %; Shimmer %; NHR; CPP; CPPS |
| Uloza et al. (2015) | 118 (84 pathological voices; 34 typical voices) | Sustained vowel | Simultaneous recordings with all devices | Studio microphone; Samsung Galaxy Note 3 | | | Mean F0; Jitter %; Shimmer %; NNE; HNR; SNR |
| Grillo et al. (2016) | 10 speakers | Sustained vowel and sentence | Simultaneous recordings with all devices | Head-mounted microphone; iPhone 5 and 6s; Samsung Galaxy S5 | | | F0 (mean and SD); Jitter %; Shimmer %; NHR; CPP; AVQI |
| Manfredi et al. (2017) | Synthesized voices | Sustained vowel | Devices recorded simultaneously through loudspeaker | Studio microphone; HTC One; Wiko Smart2 | | | Mean F0; Jitter %; Shimmer %; NHR |
| Maryn et al. | 50 (38 | Sustained vowel and | Devices recorded individually | Studio microphone; iPad 2; Google Nexus 9; iPhone | | | Median F0; Jitter (% and RAP); Shimmer (% and dB); HNR; GNE; CPPs |

| (2017) | pathological voices; 12 vocally healthy voices) | reading passage | through loudspeaker | 5s; Samsung Galaxy s5; Nokia Lumia 520 | | | |
|---|---|---|---|---|---|---|---|
| Kojima et al. (2018) | 6 vocally healthy voices | Sustained vowel | Simultaneous recordings with all devices | Studio microphone; MediaPad M3 | | | Mean F0; SNR |
| Jannetts et al. (2019) | 22 vocally healthy speakers | Sustained vowels and reading passage | Simultaneous recordings with all devices | iMac with Neumann U89i microphone; Samsung Galaxy S8 +; iPhone 6s; iPhone 7; Samsung Galaxy J3; iPhone SE | | .wav | Mean F0, CPPS, Jitter (RAP); Shimmer % |
| Freeman & De Decker (2021a) | 2 | Word list | Simultaneous recordings with all devices | PC; Mac; iPad; iPhone; Android; Zoom H4N recorder | Audacity; System; Voice Recorder; Voice Memos; Google Recorder | wav; m4a | F1 (midpoint of vowel; 35% through vowel); F2 (midpoint of vowel; 35% through vowel); A1-P0 (1/3 and 2/3 time point through vowel) |
| Freeman & De Decker (2021b) | 2 | Word list | Simultaneous recordings with all devices; recorded audio played through Praat | Zoom H4N recorder; iPad Air | Voice Memos app; Zoom; Skype; Teams; Praat to play back recordings internally | wav; m4a; mp4 | F1 (midpoint of vowel; 35% through vowel); F2 (midpoint of vowel; 35% through vowel); A1-P0 (1/3 and 2/3 time point through vowel) |

| Ge et al. (2021) | 1 | Mandarin CV syllables with four lexical tones | Simultaneous recordings with all devices | Zoom H2N recorder; iPhone 6s; iPhone 8p; Samsung Galaxy A50s; Oneplus Nord N10; DELL XPS15 (Zoom local); MacBook Pro (Zoom cloud) | built-in recorder; Smarter Recorder; Zoom | wav, m4a, aac | F0 (ten equidistant points); F1 (fifth point out of ten during the vowel portion); F2 (fifth point out of ten during the vowel portion); H1*-H2*; Jitter; Shimmer; Spectral moments (CoG, spectral skewness, kurtosis, and standard deviation) |
| Penney et al. (2021) | 4 | hVd words; read passage (only passage analyzed) | Simultaneous recordings with all devices | Zoom H6N recorder with headset microphone; MacBook Pro; Dell Mobile Precision 7530; iPhone SE; Samsung Galaxy S8 | Rode Reporter; online recorder (https://mmig.github.io/speech-to-flac/) | wav | F0; F1; F2; H1; H2; H4; H2kHz; H5kHz; H1-H2; H2-H4; H4-H2kHz; H2kHz-H5kHz; CPP; HNR05; HNR15; HNR25; HNR35: all averaged over middle third of vowel |
| Sanker et al. (2021) | 3 | Word list in carrier sentence | Study 1: Simultaneous recordings with all devices Study 2: Recordings played as input | Zoom H4N recorder; iPad; Macbook Pro; Macbook Pro with headset microphone; LG Android phone; iPhone | Zoom; Skype; Facebook messenger through Audacity; CleanFeed; Audacity | Study 1: wav; m4a Study 2: wav; mp4 | Duration/timing; F1 (mean of vowel); F2 (mean of vowel); F3 (mean); CoG (mean of consonant); Jitter (mean of vowel); F0 (mean of vowel); F0 peak alignment (in vowel); H1-H2; HNR; intensity (mean) |
| Zhang et al. (2021) | 7 | Sustained vowels | Simultaneous recordings with all devices | Zoom H6N recorder with headset microphone; smartphones (Android, iPhone, Ubuntu phones); laptop computers (Windows) | AVR app; Recorder app (for Ubuntu phone); Zoom | wav; m4a | F0; F1, F2, F3: all averaged over whole segment |

| Fahed et al. (2022) | 43 (37 typical voices, 6 with Huntington's disease) | Sustained vowel; read passage; syllable repetition task | Simultaneous recordings with all devices | Smartphone (Google Pixel 4), tablet (Samsung Tab S6 Lite), sound interface connected to a laptop with a headset microphone (6066,PDA, Denmark) | Mix Pre3 II, Sound Devices, USA | wav | F0 (mean and SD); HNR; Jitter (local and RAP); Five-point period perturbation quotient; Difference of differences of amplitude and periods; Shimmer; Amplitude perturbation quotients (APQ3, APQ5, and APQ11) |
| --- | --- | --- | --- | --- | --- | --- | --- |
| Penney et al. (2022) | 24 | hVd words read in carrier sentence | Simultaneous recordings with both devices | Neumann TL103 condenser microphone; iPhone; Samsung phones; Google phone | Audacity, Appen Research app | wav | F1, F2: temporal midpoint of each vowel |

**Key for abbreviations used in the table:**

A1 = amplitude of the highest harmonic near F1; APQ# (e.g., APQ3) = #-point amplitude perturbation quotient; AVQI = Acoustic Voice Quality Index; CoG = center of gravity; CPP = cepstral peak prominence; CPPS = smoothed cepstral peak prominence; GNE = glottal-to-noise excitation ratio; H# (e.g., H1) = harmonic #; HNR = harmonics-to-noise ratio; HNR## (e.g., HNR05) = harmonics-to-noise ratio between 0 and ## hertz (e.g., between 0 and 500 Hz); NHR = noise-to-harmonics ratio; NNE = normalized noise energy; P0 = amplitude of a low frequency harmonic peak (usually H1 or H2); RAP = relative average perturbation; SD = standard deviation; SNR = signal-to-noise ratio

**Supplementary Material 1:**

Recording instructions

"Supplementary Material 1 - Recording protocal.pdf"

Also available at: https://osf.io/cx28b

**Supplementary Material 2:**

"Supplementary Material 2 - Data & Analysis.zip"

Also available at: https://osf.io/34m5s/?view_only=9fcd872c9bf04ab69b0a2df2c3b5fb29

**Supplementary Material 3:**

"Supplementary Material 3 - Vowel_f0_results.pdf"

Also available at: https://osf.io/8749q

**Conflict of interests:**

The authors declare that they have no competing interests.

**CRediT author statement:**

**Cong Zhang**: Conceptualization, Methodology, Formal analysis, Investigation, Resources, Data Curation, Writing – Original Draft, Writing – Review & Editing, Visualization, Project administration

**Kathleen Jepson**: Conceptualization, Methodology, Formal analysis, Investigation, Resources, Writing – Original Draft, Writing – Review & Editing, Visualization

**Yu-Ying Chuang**: Formal analysis, Writing – Original Draft